%% file: paper_2_110918_arxiv.tex
\documentclass{imamat}

\usepackage[utf8]{inputenc}
\usepackage{siunitx}
\usepackage{todonotes}
\usepackage{placeins}
\usepackage{multirow}

\usetikzlibrary{shapes.misc}
\usetikzlibrary{external}
\tikzsetexternalprefix{figures/tikz/}
\definecolor{periodicsinkcolour}{RGB}{54,121,255}
\definecolor{randsinkcolour}{RGB}{255,139,61}

\tikzset{cross/.style={cross out, draw=black, minimum size=2*(#1-\pgflinewidth),
inner sep=0pt, outer sep=0pt, line width=0.65pt}, cross/.default={4pt}}

\tikzset{mycirc/.style={circle, draw=black, minimum size=2*(#1-\pgflinewidth),
inner sep=0pt, outer sep=0pt, line width=0.65pt}, mycirc/.default={3pt}}

\input{preamble.tex}

\begin{document}

\title{Homogenization approximations for unidirectional transport past randomly
distributed sinks}

\author{%
\name{Matthew J.\ Russell}
\address{School of Mathematical Sciences, The University of Nottingham, Nottingham NG7 2RD, UK} 
\name{Oliver E.\ Jensen}
\address{School of Mathematics, The University of Manchester, Manchester M13 9PL, UK}
}

\date{\today}

\maketitle

\begin{abstract}
    {
        Transport in biological systems often occurs in complex spatial 
        environments involving random structures. Motivated by such
        applications, we investigate an idealised model for solute transport
        past an array of point sinks, randomly distributed along a line, which
        remove solute via first-order kinetics. Random sink locations give rise
        to long-range spatial correlations in the solute field and influence the
        mean concentration. We present a non-standard approach to evaluating these features
        based on rationally approximating integrals of a suitable Green's
        function, which accommodates contributions varying on short and long
        lengthscales and has deterministic and stochastic components.   We
        refine the results of classical two-scale methods for a periodic sink array
        (giving more accurate higher-order corrections with non-local
        contributions) and find explicit predictions for the fluctuations in
        concentration and disorder-induced corrections to the mean for both
        weakly and strongly disordered sink locations.  Our predictions are validated across a
        large region of parameter space.
    }
    {
        homogenization; transport; spatial disorder
    }
\end{abstract}

\section{Introduction}
Spatial disorder is intrinsic to many natural systems. In biomedical
applications, for example, disorder needs careful consideration in developing
constitutive models of heterogeneous multicellular tissues and it can influence
transport processes in geometrically complex exchange organs such as the
placenta and the lung. In practice, there may be only limited knowledge of the
detailed structure of a tissue or organ of a given individual, which may simply
be expressed in terms of statistics retrieved from a population of other individuals.
If they are to support sound decision-making, theoretical models of transport or
biomechanical function should account for such uncertainties, so that
predictions can quantify the variability of outcomes within and between
individuals. Geophysical applications, for example involving transport in
subsurface porous media, raise many similar questions.

For materials with a multiscale structure (cells within a tissue, functional
exchange units within an organ, pores within a rock, etc.), homogenization
provides a powerful analytical tool, exemplified by the reduction of Stokes
equations to Darcy's equation for porous medium transport \citep{burridgekeller1981,rubinsteintorquato1989}. For a strictly
periodic medium, a variety of approaches (particularly asymptotic two-scale
convergence and spatial averaging \citep{pavliotisstuart2008,davit2013homogenization}) yield
leading-order approximations for the slowly-varying component of spatial fields,
having well-studied convergence properties. Spatial fields for such problems are
generally characterised by an almost-periodic variation at the microscale,
modulated by slow variation over much longer lengthscales. Typically a unit cell
problem at the microscale needs to be solved (or averaged) in order to derive an
approximation capturing the macroscale variation. Such approaches can be
extended to account for smooth variation in the properties of the unit cells
over long lengthscales \citep{davit2013homogenization, bruna2015diffusion}. For
materials with a random microstructure that lacks exact periodicity, however,
less is understood about the accuracy of (deterministic) homogenization
approximations, and leading-order approximations generally capture only mean
behaviour. For proper uncertainty quantification, it is necessary to determine
corrections that capture the stochastic variability in the quantity of interest.
Such corrections are likely to be entangled with the discrete-to-continuous
corrections of the classical homogenization approximation and are typically
nonlocal \citep{heitzinger2014, lebris2014, wood2013}.

Maternal circulation in the placenta provides an interesting test-bed for such
ideas \citep{jensenchernyavsky2019}. This organ exchanges dissolved gases, nutrients and other materials
between maternal and fetal blood. Fetal blood vessels are arranged in tree-like
structures called villi; maternal blood
in the intervillous space flows past the outside of their branches, effectively
passing through a disordered porous medium. Initial models described maternal
blood flow using Darcy’s equation (with a uniform permeability) in two spatial
dimensions \citep{erian1977maternal}, with solute transport being described by a
slowly-varying (macroscale) advection/uptake equation with a uniform uptake coefficient
\citep{chernyavsky2010mathematical}. In an effort to understand the role of
disorder in the arrangement of fetal vessels,
\cite{chernyavsky2011transport,chernyavsky2012characterizing} studied simplified
models of solute transport, reducing villous branches to point sinks
(distributed along a line according to a prescribed distribution) and describing
solute transport between sinks using a one-dimensional advection-diffusion
equation. These studies tested the quality of the macroscale approximation in relation to simulations of the solute
concentration under different sink realisations. Direct evaluation of the error
(the homogenization residue) revealed its varying character across parameter
space, its long-range covariance structure and its dependence on the statistical properties
of the underlying sink distribution. In particular, parameter regimes were
identified where the macroscale approximation fails because stochastic
sink-to-sink variations in the solute concentration become dominant.

\cite{chernyavsky2011transport,chernyavsky2012characterizing} used an algebraic
method to compute the homogenization residue directly, for zeroth-order uptake
kinetics, which worked only over a limited range of parameter space. A more
robust approach was presented by \cite{russell2016stochastic}, in a related
problem assuming first-order kinetics and variable sink strength (rather than
sink location). When disorder is weak, an expansion can be developed in which a
deterministic periodic problem at leading order (which is readily homogenized)
is perturbed to give a stochastic linear problem at the following order.
Linearity allows the disorder due to individual sinks to be evaluated
independently using a Green’s function; the individual contributions are then
assembled (exploiting the central limit theorem) to capture the overall disorder
in the system, which has an inherently nonlocal structure. This method does not
suffer the parameter-space restriction of earlier approaches, and it is
developed further below. Taking the expansion to higher order,
\cite{russell2016stochastic} demonstrated how the macroscale approximation
has a small but systematic error in the presence of weak disorder. The value of Green's
functions in evaluating corrector fields was demonstrated also by \cite{wood2013} and \cite{heitzinger2014}; 
the latter authors for example considered a Poisson problem with a distributed source that is statistically uncorrelated 
over a periodic array of cells.

In the present study we consider how the random spatial location (along a line)
of identical first-order sinks influences the distribution of a solute that
moves between them by advection and diffusion. We consider periodic, weakly
disordered and strongly disordered sink locations. In the periodic case, we use
a Green's function approach (instead of the traditional two-scale expansion,
which relies on an \textit{ad hoc} periodicity assumption) to derive the
macroscale solution and its corrections. The methods deviate in their
predictions at sub-leading-order: we demonstrate numerically that the Green's
function approach is more accurate than the classical approach. To address
disorder, we construct an empirical expansion about the macroscale solution,
again exploiting Green's functions, correcting successively for the
discrete-to-continuous and periodic-to-disordered effects. The expansion is
shown to be effective both in the weakly disordered limit (as demonstrated in
Russell et al. 2016) and the strongly disordered case (when sinks are
distributed uniformly randomly in a finite interval). We do not attempt to
provide rigorous convergence proofs; however having adopted a non-standard
expansion, we apply careful asymptotic techniques in order to approximate the
sums and integrals that arise and to establish their relative magnitudes.

\section{Model definition}
\begin{figure}[t!]
    \centering
    \includegraphics{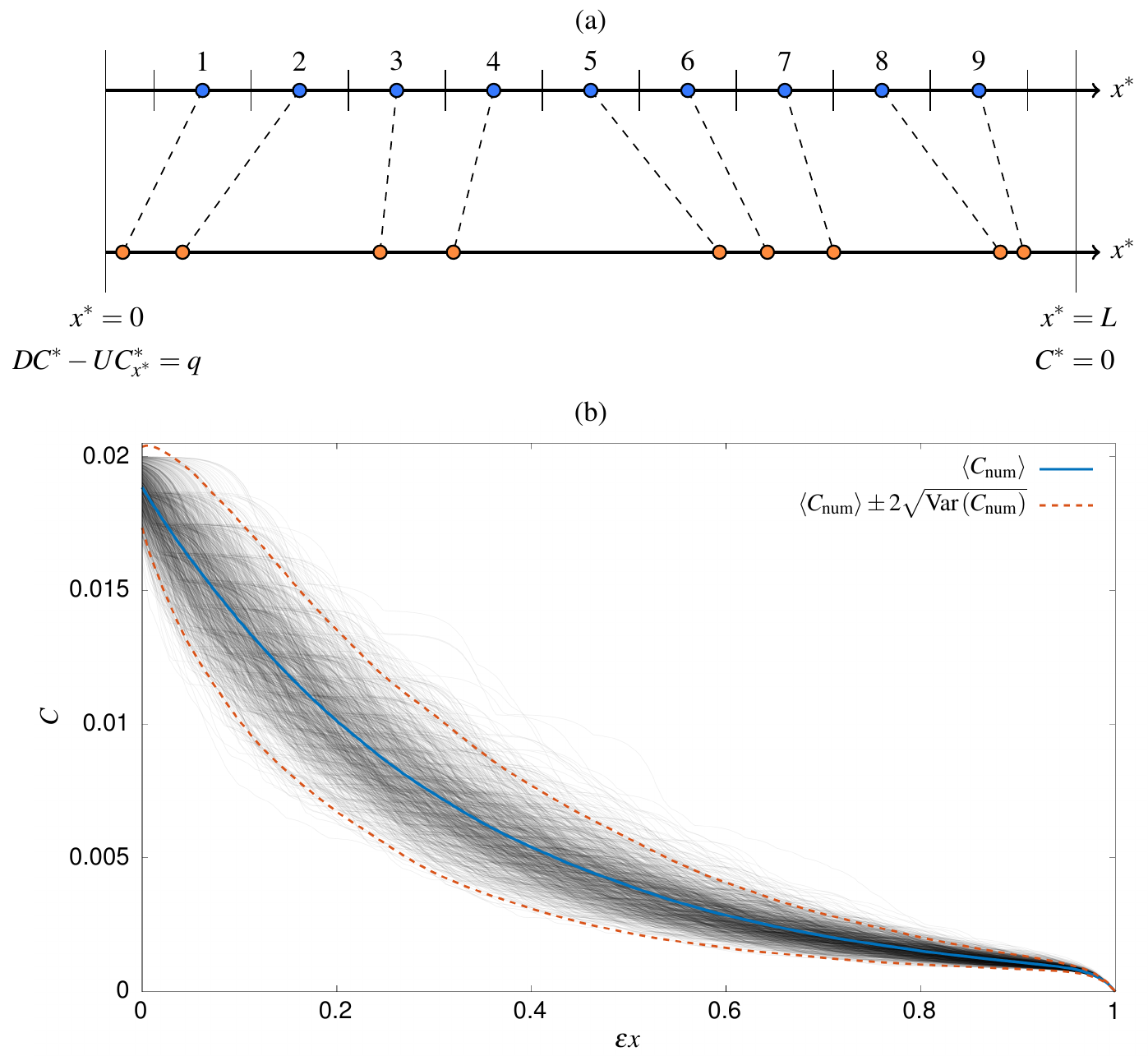}
    \caption{(a) Diagram of the problem domain, with inlet and outlet conditions
        indicated below their respective boundaries. On the top axis, a periodic
        configuration of $N = 9$ sinks is shown. Blue circles mark the sink
        locations and vertical lines delimit the associated unit cells. The
        bottom axis shows one realisation of a random distribution of sinks,
        denoted by orange circles. The dashed lines indicate how we label sinks
        in ascending order, regardless of how they were drawn from the
        distribution. (b) An ensemble of $10^3$ concentration profiles (thin
        black lines). Each realisation has $N=49$ sinks (\ie $\ep=0.02$) with
        uniformly random locations, and $\pe=1$, $\da=\frac{1}{2}\ep^{1/2}$.
        Also shown is the sample mean (solid blue) and two standard deviations
        either side of the mean (dashed orange).
    }
    \label{fig:domain+realisations}
\end{figure}

We model steady transport of a solute past a linear array of point sinks using
an advection-diffusion-uptake equation. The sinks have first-order uptake
kinetics and sit at discrete locations $x^* = \xi_j^*$, $j = 1,\dotsc,N$; see
Figure~\ref{fig:domain+realisations}(a). We introduce the sink density, P\'eclet
number, Damk\"ohler number and concentration scale as
\begin{equation}
    \label{eqn:params}
    \ep = \frac{\ell}{L} = \frac{1}{N+1}, \quad \pe = \frac{U\ell}{D}, \quad
    \da = \frac{S_0 \ell}{D}, \quad C_0 = \frac{qL}{D},
\end{equation}
respectively, where $\ell$ is the average inter-sink distance, $L$ is the domain
length, $U$ is the advection velocity, $D$ is the diffusion coefficient, $S_0$
is the sink strength per unit concentration and $q$ is the flux at the inlet
boundary. At the downstream boundary we impose zero concentration, $C^*|_{x^*=L}
= 0$. The P\'eclet number $\pe$ represents the strength of advection relative to
diffusion and the Damk\"ohler number $\da$ represents the strength of uptake
relative to diffusion. The governing equation and boundary conditions for the
solute concentration $C^*(x^*)$ are
\begin{subequations}
    \label{eqn:adre_var_loc_dim}
    \begin{gather}
        D C^*_{x^* x^*} - U C^*_{x^*} = S_0 C^*(x^*)\sum_{j=1}^N
        \delta(x^*-\xi_j^*), \quad 0 \le x^* \le L,\\
        U C^*|_{x^*=0} - D C^*_{x^*}|_{x^*=0} = q, \quad C^*|_{x^*=L} = 0.
    \end{gather}
\end{subequations}
Introducing the dimensionless variables
\begin{equation}
    \label{eqn:scalings}
    x = x^*/\ell, \quad \xi_j = \xi_j^*/\ell, \quad C(x) = C^*(x^*)/C_0,
\end{equation}
\eqref{eqn:adre_var_loc_dim} becomes,
\begin{subequations}
    \label{eqn:adre_var_loc}
    \begin{gather}
        \label{eqn:adre_var_loc_eqn}
        C_{xx} - \pe C_x = \da C(x) \sum_{j=1}^N \delta(x - \xi_j), \quad 0 \le
        x \le \epm,\\
        \label{eqn:adre_var_loc_bcs}
        \pe C|_0 - C_x|_0 = \ep, \quad C|_\epm = 0.
    \end{gather}
\end{subequations}
For later convenience, we set $\xi_0 = 0$ and $\xi_{N+1} = \epm$. Integrating over the whole domain yields the overall flux balance
\begin{equation}
    \ep+C_x |_\epm = \da \sum_{j=1}^N C(\xi_j)
    \label{eq:netflux}
\end{equation}
which provides a direct method of determining the net uptake by the internal sinks. We will
consider the following sink distributions: periodically-located, $\xi_j = j$;
normally-perturbed from a periodic arrangement, $\xi_j \sim \mathcal{N}(
j,\sigma^2)$ for some small variance $\sigma^2$; and uniformly distributed in
the domain according to $\mathcal{U}(0,\epm)$, sorted into ascending order. 
In the uniformly-random case, the sink locations $\xi_j$ are order statistics of the
uniform distribution and are then spatially correlated, unlike the
independently drawn uniform variates.

The problem described by \eqref{eqn:adre_var_loc} involves a number of spatial
scales (\eg domain length and sink-to-sink distance) over which advection,
diffusion and uptake take place. Spatial disorder, in the form of random
sink locations, adds additional complexity to the problem. These features are
illustrated in Figure~\ref{fig:domain+realisations}(b), which shows $10^3$
concentration profiles generated numerically using the method described in
Appendix~\ref{app:numerics}. Each realisation has $N=49$ sinks whose locations
are uniformly randomly distributed across the domain. The parameters are
$\pe=1$, $\da=\tfrac{1}{2}\ep^{1/2}$, which is a regime with strong advection
and uptake characterised by a prominent sink-to-sink ``staircase'' structure and
uptake across a large portion of the domain.  We seek to characterise the mean and (co)variance of the concentration across the domain.

\section{Constructing an expansion}

We introduce the linear differential operator $\lx = \partial_x^2 - \pe
\partial_x - \da$ and boundary conditions $\bx C = \{(\pe - \partial_x)C|_0,
C|_\epm\}$, and let $\Ch(x)$ satisfy the homogenized analogue of
\eqref{eqn:adre_var_loc}, in which point sinks are replaced by a smoothly varying term,
\begin{equation}
    \label{eqn:C_H_eqn_bcs}
    \lx \Ch = 0,\quad \bx \Ch = \{\ep,0\}, \quad 0 \le x \le \epm.
\end{equation}
Defining $\phi \equiv \sqrt{\da + \pe^2/4}$ and $g(x) \equiv \pe\sinh(\phi x) +
2\phi\cosh(\phi x)$, \eqref{eqn:C_H_eqn_bcs} has the exact solution
\begin{equation}
    \label{eqn:C_H}
    \Ch(x) = \frac{2\ep}{g(\epm)}e^{\frac{1}{2}{\pe x}}\sinh(\phi[\epm - x]),
    \quad 0 \le x \le \epm.
\end{equation}
$\Ch$ represents the leading-order homogenized solution to
\eqref{eqn:adre_var_loc} as $\ep \to 0$ for $\pe = \bo{\ep}$, $\da =
\bo{\ep^2}$, when the sinks are distributed periodically; see
\cite{russell2016stochastic} and Appendix~\ref{app:periodic_homog}, where we
revisit the classical two-scale expansion for this problem. We will initially
work in this parameter regime, for which there is a dominant balance between
advection, diffusion and uptake across the domain. However, unlike the classical
approach, we make no assumption in what follows about $C$ depending on
independent long- and short-range variables.

The Green's function $G(x,y)$ associated with $\lx$ under homogeneous boundary
conditions $\bx G = \{0,0\}$ satisfies $\lx G = \delta(x-y)$. We write
\begin{equation}
    \label{eqn:G_piecewise}
    G(x,y) =
    \begin{cases}
        G^-(x,y), & 0 \le x \le y \le \epm,\\
        G^+(x,y), & 0 \le y \le x \le \epm.
    \end{cases}
\end{equation}
$G(x,y)$ is piecewise smooth, continuous at $x = y$ and satisfies the following jump
conditions, resulting from the point source at $x = y$:
\begin{equation}
    \label{eqn:G_jump_cond}
    G^+_x(y,y) - G^-_x(y,y) = 1, \quad G^-_y(x,x) - G^+_y(x,x) = 1.
\end{equation}
The two pieces of the Green's function can be expressed as
\begin{equation}
    \label{eqn:G}
    \begin{aligned}
        G^-(x,y) &= \frac{g(x)}{\phi
        g(\epm)}e^{\frac{1}{2}{\pe}(x-y)}\sinh(\phi[y-\epm]),\\
        G^+(x,y) &= e^{\pe(x-y)}G^-(y,x).
    \end{aligned}
\end{equation}
Later we will use the identity
\begin{equation}
    \label{eqn:C_H_G+}
    \Ch(x) = -\ep G^+(x,0).
\end{equation}

\begin{figure}[t!]
    \centering
    \includegraphics[width=\textwidth]{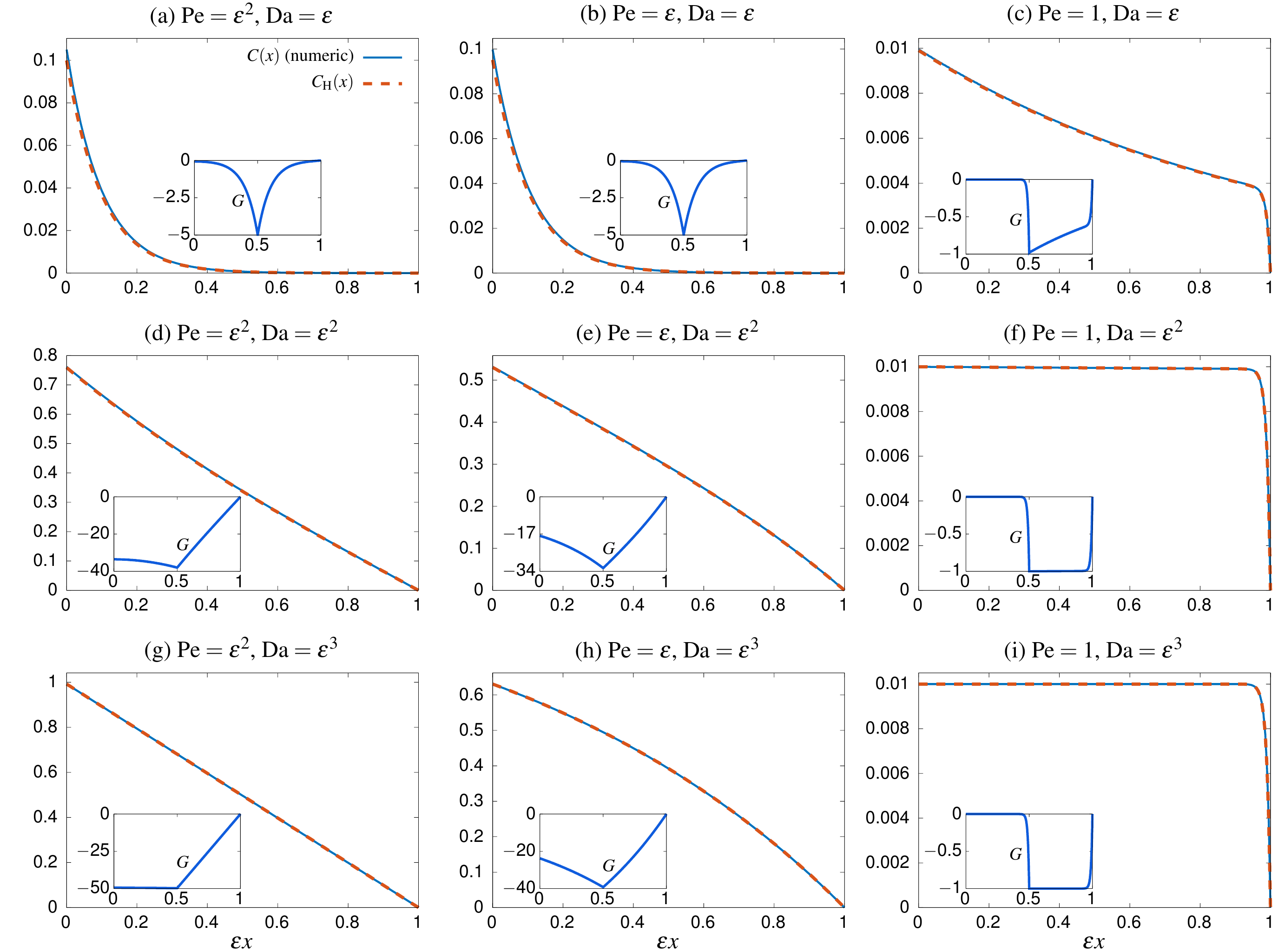}
    \caption{Comparison between the numerical solution of
        \eqref{eqn:adre_var_loc} $C(x)$ with a periodic sink array (solid blue)
        and the leading-order homogenized approximation $\Ch$ satisfying
        \eqref{eqn:C_H} (dashed orange), both plotted versus $\ep x$. Insets
        show the corresponding Green's function \eqref{eqn:G_piecewise} with the
        point source located in the centre of the domain, $y=\tfrac{1}{2}\epm$
        for illustrative purposes. There are $N=99$ sinks in all cases, so $\ep
        = 10^{-2}$, and the $\pe$, $\da$ values are indicated in the panel
        titles, with all combinations of $\pe \in \{\ep^2, \ep, 1\}$, $\da \in
        \{\ep^3, \ep^2, \ep\}$ shown.
    }
    \label{fig:C_vs_C_H_periodic}
\end{figure}

Figure~\ref{fig:C_vs_C_H_periodic} compares the leading-order approximation
$\Ch$ with numerical solutions of \eqref{eqn:adre_var_loc} (obtained using the
method described in Appendix~\ref{app:numerics} for $N=99$ periodically located
sinks and a range of $(\pe$, $\da)$-values, showing good agreement.  Also shown as insets in each panel
are illustrative plots of the Green's function \eqref{eqn:G_piecewise}.

Inspection of \eqref{eqn:C_H} and \eqref{eqn:G} reveals that $\Ch(x)$ and $\ep
G(x,y)$ vary by $\bo{1}$ as $x$, $y$ vary across the domain (that is as $\ep x$,
$\ep y$ vary by $\bo{1}$), in the distinguished limit $\pe = \bo{\ep}$, $\da =
\bo{\ep^2}$. This can be seen in Figure~\ref{fig:C_vs_C_H_periodic}(e).
Increasing uptake relative to diffusion, characterised by elevated $\da$, leads
to more rapid decay near the inlet (see Figure~\ref{fig:C_vs_C_H_periodic}b);
increasing advection relative to diffusion, characterised by elevated $\pe$,
leads to a diffusive boundary layer near the outlet and, for $G$, near $x=y$ (see
Figure~\ref{fig:C_vs_C_H_periodic}f). We introduce the notation 
\begin{equation}
    \label{eqn:gch_def}
    \gch|_{x,y} \equiv G(x,y)\Ch(y)
\end{equation}
and use the corresponding notation with $G^+$ and $G^-$ in place of $G$, noting
that for $\pe=\bo{\ep}$, $\da=\bo{\ep^2}$, each derivative of $\Ch$ and $G$ with
respect to $x$ or $y$ reduces its magnitude by $\bo{\ep}$.

To account for the discrete nature of the sinks and their spatial distribution
in the domain, we pose an expansion for solutions to \eqref{eqn:adre_var_loc}
consisting of the approximation \eqref{eqn:C_H} and a series of correction
terms. We therefore write $C = \Ch + \hat C$, where the corrections $\hat C$
satisfy $\bx \hat C = \{0,0\}$ and
\begin{equation}
    \label{eqn:C_hat}
    \begin{aligned}
        \lx \hat C
        = \da \Biggl\{&\Ch(x)\Bigl[\sum\nolimits_{j=1}^N \delta(x-j) - 1\Bigr]
            + \Ch(x) \sum\nolimits_{j=1}^N \Bigl[\delta(x-\xi_j) - \delta(x-j)\Bigr]\\
            +&\hat C(x)\Bigl[\sum\nolimits_{j=1}^N \delta(x-j) - 1\Bigr]
            + \hat C(x) \sum\nolimits_{j=1}^N \Bigl[\delta(x-\xi_j) - \delta(x-j)\Bigr]
        \Biggr\}.
    \end{aligned}
\end{equation}
Since \eqref{eqn:C_hat} is linear, we may consider the solution of the
sub-problems associated with each right-hand sum separately. The first is
\begin{equation}
    \label{eqn:C_a_eqn}
    \lx \hat C_a = \da \Ch(x)\Biggl[\sum_{j=1}^N \delta(x-j) - 1\Biggr],
    \quad \bx \hat C_a = \{0,0\},
\end{equation}
which describes a transport problem modelling the difference between a periodic
sink arrangement and a smooth sink distribution, with strengths modulated by the
leading-order concentration profile $\Ch$. Using the Green's function
\eqref{eqn:G_piecewise} we can write the solution as
\begin{equation}
    \label{eqn:C_a_int}
    \hat C_a(x) = \da \int_0^\epm \gch|_{x,y}
    \Biggl[\sum_{j=1}^N \delta(y-j) - 1\Biggr] \dd y.
\end{equation}
We expect $\hat C_a$ to provide the dominant corrections due to discrete-sink
effects to the leading-order solution $\Ch$. Similarly, the second
sub-problem from \eqref{eqn:C_hat} is
\begin{equation}
    \label{eqn:C_b_eqn}
    \lx \hat C_b = \da \Ch(x)\sum_{j=1}^N \left[\delta(x-\xi_j) -
    \delta(x-j)\right], \quad \bx \hat C_b = \{0,0\},
\end{equation}
so that
\begin{equation}
    \label{eqn:C_b_int}
    \begin{aligned}
        \hat C_b(x) &= \da \int_0^\epm \gch|_{x,y}
        \sum_{j=1}^N\left[\delta(y-\xi_j) - \delta(y-j)\right] \dd y\\
        &= \da \sum_{j=1}^N \left[\gch|_{x,\xi_j} - \gch|_{x,j}\right],
    \end{aligned}
\end{equation}
which captures the effects of displacing sinks from a periodic to a disordered
arrangement, with strengths again given by $\Ch$. We may recursively continue
the expansion in the following way
\begin{equation}
    \label{eqn:C_expansion}
    C = \Ch + (\hat C_a + \hat C_b) + (\hat C_{aa} + \hat C_{ab} + \hat C_{ba} +
    \hat C_{bb}) + \dotsb,
\end{equation}
(a form of Duhamel expansion \citep{bal2011}) where the remaining subproblems in
\eqref{eqn:C_hat} become
\begin{equation}
    \label{eqn:C_xx_eqns}
    \begin{aligned}
        \lx \hat C_{aa} &= \da \hat C_a \Biggl[\sum_{j=1}^N \delta(x-j) - 1
        \Biggr], \quad
        \lx \hat C_{ab} = \da \hat C_a \sum_{j=1}^N \left[\delta(x-\xi_j) -
        \delta(x-j)\right]\\
        \lx \hat C_{ba} &= \da \hat C_b \Biggl[\sum_{j=1}^N \delta(x-j) - 1
        \Biggr], \quad
        \lx \hat C_{bb} = \da \hat C_b \sum_{j=1}^N \left[\delta(x-\xi_j) -
        \delta(x-j)\right],
    \end{aligned}
\end{equation}
etc., each with homogeneous boundary conditions. Each term with a subscript
containing one or more $b$ involves a random variable. The problem of a
deterministic, periodic sink array is fully described by $\Ch$, $\hat C_a$,
$\hat C_{aa}$, \dots. This series is not assumed to be asymptotic in general,
but the size of each term can be estimated \textit{a posteriori} based on their
dependence on $\Ch$,  $G$ and the parameters $\da$ and $\pe$ to suggest an asymptotic reordering of terms and to
assess convergence. Magnitudes of the remaining terms will depend on the choice
of sink distribution so we must analyse each case separately.  

The following strategy will be used to estimate magnitudes in terms of $\ep$ in
the distinguished limit $\pe = \bo{\ep}$, $\da = \bo{\ep^2}$, for which $\Ch$ and $G$ are piecewise smooth on the macroscale (Figure~\ref{fig:C_vs_C_H_periodic}e). In this limit sums, perhaps with
finitely many terms excluded, and integrals over the domain will contribute a
factor of $\epm \approx N$; $G$ contributes a factor of $\epm$ and $\Ch$ a
factor of $1$; each successive derivative of $G$ and $\Ch$ with respect to $x$
or $y$ gains an additional factor $\ep$ in magnitude, reflecting the slow
variation of these functions across the domain (see
Figure~\ref{fig:C_vs_C_H_periodic}e). Alternative scaling arguments will be required in other parameter regimes.

\section{The periodic problem}
\label{sec:periodic}

\subsection{Solving for $\hat C_a$}
To solve \eqref{eqn:C_a_eqn} for $\hat C_a$, we split the domain $[0, \epm]$
into unit cells $(j-\frac{1}{2}, j+\frac{1}{2})$ for $j = 1,2,\dotsc,N$, and two
half-cells $[0,\frac{1}{2})$ and $(\epm-\frac{1}{2}, \epm]$ at the ends. The
cell which contains $x$, say $j = k$ where $k \equiv \floor{x+\frac{1}{2}}$, is
treated separately and we split the integral at $y = x$ to allow for careful
handling of discontinuities. We Taylor expand $\gch|_{x,y}$ inside the integral
around $y = j$ for $j \neq k$, around $y = \pm x$ for $j = k$, and around $y =
0$ and $y = \epm$ for the inlet and outlet half-cells, respectively. Then,
\eqref{eqn:C_a_int} becomes
\begin{equation}
    \label{eqn:C_a_expanded}
    \begin{aligned}
        \hat C_a(x) = \da\Biggl\{
            &\sum_{j \neq k} \int_{j-1/2}^{j+1/2}
            \left[\gch|_{x,j} + (y-j)\gch_y|_{x,j} +
            \tfrac{1}{2}(y-j)^2\gch_{yy}|_{x,j} + \dotsb \right]
            [\delta(y-j) - 1] \dd y\\
            +&\int_{k-1/2}^x
            \left[\gpch|_{x,x} + (y-x)\gpch_y|_{x,x} +
            \tfrac{1}{2}(y-x)^2\gpch_{yy}|_{x,x} + \dotsb \right]
            [\delta(y-k) - 1] \dd y\\
            +&\int_x^{k+1/2}
            \left[\gmch|_{x,x} + (y-x)\gmch_y|_{x,x} +
            \tfrac{1}{2}(y-x)^2\gmch_{yy}|_{x,x} + \dotsb \right]
            [\delta(y-k) - 1] \dd y\\
            -&\int_0^{1/2}
            \left[\gch|_{x,0} + y\gch_y|_{x,0} +
            \tfrac{1}{2}y^2\gch_{yy}|_{x,0} + \dotsb \right]
            \dd y\\
            -&\int_{\epm-1/2}^\epm
            \left[\gch|_{x,\epm} + (y-\epm)\gch_y|_{x,\epm} +
            \tfrac{1}{2}(y-\epm)^2\gch_{yy}|_{x,\epm} + \dotsb \right]
            \dd y
        \Biggr\}.
    \end{aligned}
\end{equation}
This expansion of $(G\Ch)$ reduces each integrand to a sum of polynomials, each
multiplied by $[\delta(x-j) - 1]$ or similar (except for the half-cell
integrals). Integrating over the two half-cells,
\begin{equation*}
    \begin{aligned}
        -\int_0^{1/2} [\cdots] \dd y &= -\tfrac{1}{2} \gpch|_{x,0} - \tfrac{1}{8}
        \gpch_y|_{x,0} - \tfrac{1}{48} \gpch_{yy}|_{x,0} + \dotsb,\\
        -\int_{\epm-1/2}^{\epm} [\dotsm] \dd y &=
        -\tfrac{1}{24} (G^-_y \Chs{y})|_{x,\epm} + \dotsb,
    \end{aligned}
\end{equation*}
where we have eliminated terms using the boundary conditions
\eqref{eqn:C_H_eqn_bcs}. The first two terms of the first integral in
\eqref{eqn:C_a_expanded} vanish (see Appendix~\ref{app:identities}), as does the
first term of the integrals over cell $j=k$, for which we also use that
$G^+|_{x,x} = G^-|_{x,x}$. We use the identities in
Appendix~\ref{app:identities} again to calculate the remaining integrals, and
\begin{equation*}
    \begin{aligned}
        \hat C_a(x) = \da
        \biggl\{
            &-\tfrac{1}{24} \sum_{j \neq k} \gch_{yy}|_{x,j}
            -\tfrac{1}{24} (G\Chs{yy})|_{x,x}\\
            &+ (G^+_y\Ch)|_{x,x} \left[(k-x)H(x-k)
            +\tfrac{1}{2}(x-k+\tfrac{1}{2})^2\right]\\
            &+ (G^-_y\Ch)|_{x,x} \left[(k-x)H(k-x) -
            \tfrac{1}{2}(x-k-\tfrac{1}{2})^2\right]\\
            &+ \tfrac{1}{2} (G^+_{yy}\Ch + 2G^+_y\Chs y)|_{x,x}
            \left[(k-x)^2H(x-k) - \tfrac{1}{3}(x-k+\tfrac{1}{2})^3\right]\\
            &+ \tfrac{1}{2} (G^-_{yy}\Ch + 2G^-_y\Chs y)|_{x,x}
            \left[(k-x)^2H(k-x) + \tfrac{1}{3}(x-k-\tfrac{1}{2})^3\right]\\
            &-\tfrac{1}{2}\gpch|_{x,0} - \tfrac{1}{8}\gpch_y|_{x,0}
            - \tfrac{1}{48}\gpch_{yy}|_{x,0} - \tfrac{1}{24}(G^-_y\Chs{y})|_{x,\epm}
            + \dotsb
        \biggr\}.
    \end{aligned}
\end{equation*}
Using the jump conditions \eqref{eqn:G_jump_cond}, we write $G^-_y$ in terms of
$G^+_y$ to give
\begin{equation}
    \label{eqn:C_a_integrated}
    \begin{aligned}
        \hat C_a(x) = \da
        \biggl\{
            &-\tfrac{1}{24} \sum_{j \neq k} \gch_{yy}|_{x,j}
            -\tfrac{1}{24} (G\Chs{yy})|_{x,x} - \tfrac{1}{12}(G^+_y\Chs{y})|_{x,x}\\
            &+ \Ch(x) \left[(k-x)H(k-x) - \tfrac{1}{2}(x-k-\tfrac{1}{2})^2\right]\\
            &+ \Chs{y}(x) \left[(k-x)^2 H(k-x) + \tfrac{1}{3}(x-k-\tfrac{1}{2})^3\right]\\
            &+ \tfrac{1}{2} (G^+_{yy}\Ch)|_{x,x}
            \left[(k-x)^2H(x-k) - \tfrac{1}{3}(x-k+\tfrac{1}{2})^3\right]\\
            &+ \tfrac{1}{2} (G^-_{yy}\Ch)|_{x,x}
            \left[(k-x)^2H(k-x) + \tfrac{1}{3}(x-k-\tfrac{1}{2})^3\right]\\
            &-\tfrac{1}{2}\gpch|_{x,0} - \tfrac{1}{8}\gpch_y|_{x,0}
            - \tfrac{1}{48}\gpch_{yy}|_{x,0} - \tfrac{1}{24}(G^-_y\Chs{y})|_{x,\epm}
            + \dotsb
        \biggr\}.
    \end{aligned}
\end{equation}
Recalling that $G = \bo{\epm}$ and $\Ch = \bo{1}$ when $\pe = \bo{\ep}$, $\da =
\bo{\ep^2}$, we can collect the terms in \eqref{eqn:C_a_integrated} by
magnitude:
\begin{equation*}
    \begin{aligned}
        \hat C_a(x) = \da
        \biggl\{
            &-\tfrac{1}{2}\gpch|_{x,0} - \tfrac{1}{8}\gpch_y|_{x,0}
            + \Ch(x)\left[(k-x) H(k-x) -
            \tfrac{1}{2}(x-k-\tfrac{1}{2})^2\right]\\
            &\quad -\tfrac{1}{24}\sum_{j \neq k}\gch_{yy}|_{x,j} + \bo{\ep}
        \biggr\}.
    \end{aligned}
\end{equation*}
Using \eqref{eqn:sum_to_int} to approximate sums with integrals and using the
jump condition \eqref{eqn:G_jump_cond}, gives
\begin{equation}
    -\tfrac{1}{24}\sum_{j \neq k} \gch_{yy}|_{x,j} = \tfrac{1}{24}\gch_y|_{x,0}
    + \tfrac{1}{24}\Ch(x) + \bo{\ep}.
\end{equation}
Defining $f(x) \equiv -xH(-x) - \tfrac{1}{2}(x-\tfrac{1}{2})^2 +
\tfrac{1}{24}$, for $|x| < 1/2$ (and zero otherwise), so that $f$ has zero unit cell average $\int_{-1/2}^{1/2}
f(x) \dd x = 0$, it follows that
\begin{equation}
    \label{eqn:C_a}
    \hat C_a(x) = \da
    \Bigl[
        \tfrac{1}{2}\epm \Ch(0)\Ch(x) + \tfrac{1}{12}\epm \Chs{y}(0)\Ch(x) +
        f(x-k)\Ch(x) + \bo{\ep}
    \Bigr],\quad k=\floor{x+\tfrac{1}{2}}.
\end{equation}
The term $f(x-k)\Ch(x)$ in \eqref{eqn:C_a} varies rapidly on the scale of
individual unit cells, modulated on an $\bo{\epm}$ lengthscale by $\Ch$. The two
additional contributions are slowly varying. In the distinguished limit, the
slowly varying terms have magnitudes $\bo{\ep}$ and $\bo{\ep^2}$ respectively and the
oscillatory term has magnitude $\bo{\ep^2}$. The neglected terms are of
magnitude $\bo{\ep^3}$; however, we show below that $\hat C_{aa}$ also
contributes at $\bo{\ep^2}$.

\subsection{Solving for $\hat C_{aa}$}
With periodically located sinks, the next non-zero
term in the expansion \eqref{eqn:C_expansion} is $\hat C_{aa}$, satisfying
\eqref{eqn:C_xx_eqns}. Recognising the recursive nature of the expansion, $\hat
C_{aa}$ takes the same form as $\hat C_a$, but with the role of $\Ch$ fulfilled
by $\hat C_a$ as follows,
\begin{equation}
    \label{eqn:C_aa_pre}
    \hat C_{aa}(x) = \da
    \Bigl[
        -\tfrac{1}{2}G^+(x,0)\hat C_a(0) - \tfrac{1}{12}\bigl(G^+(x,y)\hat
        C_a(y)\bigr)_y|_{y=0} + f(x-k)\hat C_a(x) + \dotsb
    \Bigr].
\end{equation}
Retaining only the leading-order, slowly varying terms in $\hat C_a$, gives
\begin{equation}
    \label{eqn:C_aa}
    \hat C_{aa}(x) = \da^2 \ep^{-2} \Bigl[\tfrac{1}{4}\bigl(\Ch(0)\bigr)^2\Ch(x)
        + \dotsb
    \Bigr].
\end{equation}
At sub-leading orders $\hat C_{aa}$ contains oscillatory terms from $\hat C_a$
and further terms arising from the $f(x-k)\hat C_a(x)$ term in
\eqref{eqn:C_aa_pre}.

In the distinguished limit, $\hat C_{aa}$ has magnitude $\bo{\ep^2}$ so must be
included in an expansion of the concentration up to this order. The next
correction, $\hat C_{aaa}$, is related to $\hat C_{aa}$ in the same way that
$\hat C_{aa}$ is related to $\hat C_{a}$ in \eqref{eqn:C_aa_pre}. The largest
terms in $\hat C_{aaa}$ are therefore $\bo{\ep^3}$, and since this recursive
pattern continues for the higher corrections we conclude that all contributions
up to $\bo{\ep^2}$ are captured by the corrections up to $\hat C_{aa}$.
Interestingly, our prediction of the sub-leading-order terms for the
periodically-located sinks problem contain extra terms up to order $\bo{\ep^2}$
compared with results from classical two-scale homogenization; see
Appendix~\ref{app:periodic_homog} for a derivation of the classical results.

For the purposes of comparing the theoretical predictions with simulations, we
define the following residual (using leading-order expressions for $\hat{C}_a$ and $\hat{C}_{aa}$)
\begin{subequations}
\label{eqn:resid}
\begin{align}
    r(x) &\equiv  C_\text{num}(x) - \left(\Ch(x) + \left[\hat C_a(x) - \da
    f(x-k)\Ch(x)\right] + \hat C_{aa}(x)\right) \\
    &= C_{\text{num}}(x) - \Ch(x)
    \left(
        1 + \tfrac{1}{2} \da\epm \left[ \Ch(0) + \tfrac{1}{6}{\Ch}_y(0)\right] +
        \tfrac{1}{4} \da^2 \ep^{-2} [\Ch(0)]^2
    \right),
\end{align}
\end{subequations}
in which we have subtracted from numerical solutions $C_\text{num}(x)$ of the
full problem \eqref{eqn:adre_var_loc} all the terms in $\hat C_a$ and $\hat
C_{aa}$ appearing in (\ref{eqn:C_a}, \ref{eqn:C_aa}) which vary slowly across
the domain, plus of course $\Ch$. What remains, to leading-order, is the
numerical prediction of the sink-to-sink oscillating part of the solution. We
further define a residual based on the results of the classical method in
Appendix~\ref{app:periodic_homog}, which to the same level of accuracy is
\begin{equation}
    \rc(x) \equiv C_\text{num}(x) - \Ch(x)\Bigl( 1 +
    \tfrac{1}{2}\da\epm\Ch(0) \Bigr).
\end{equation}
\begin{figure}[t!]
    \centering
    \includegraphics[width=\textwidth]{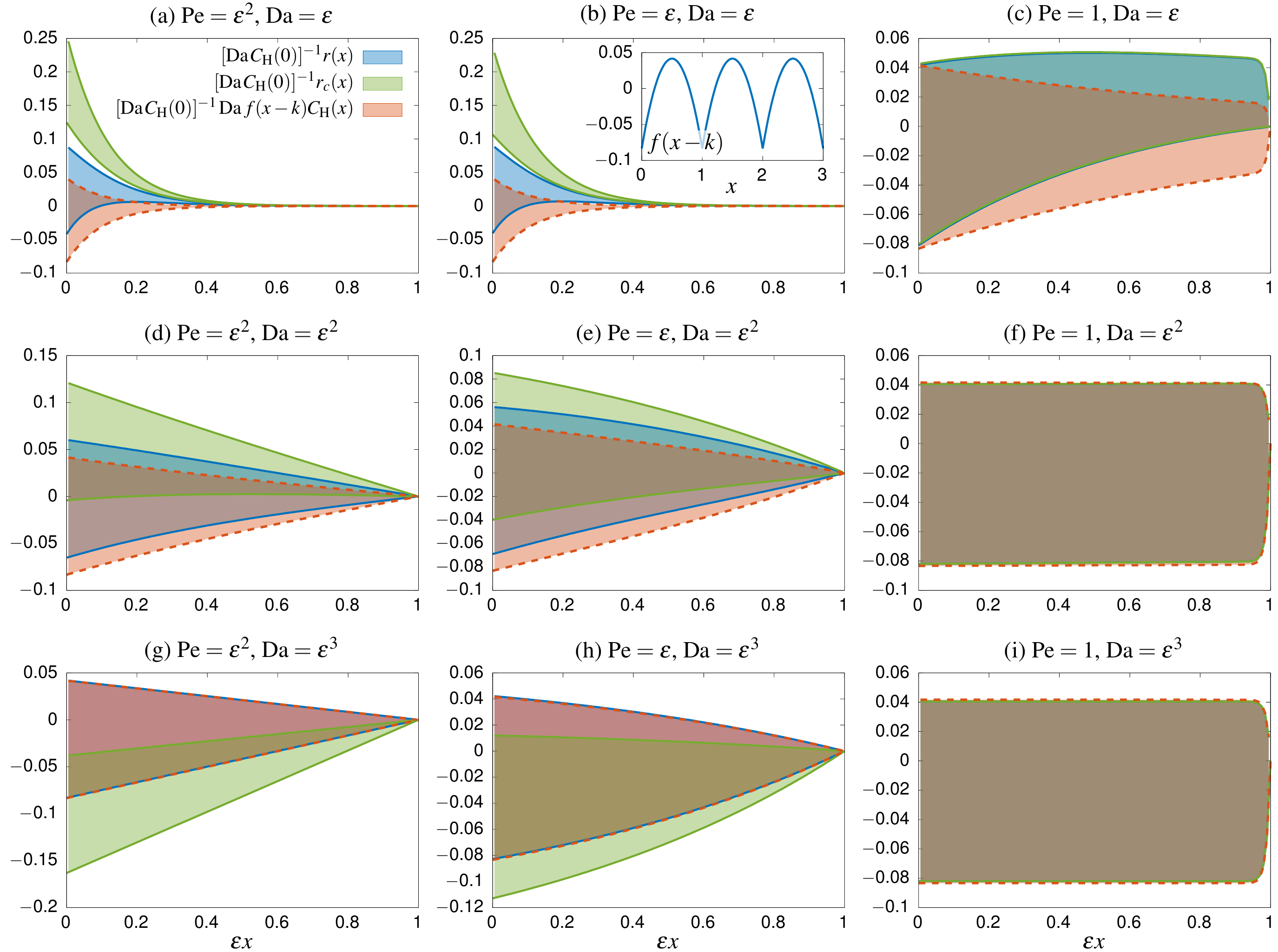}
    \caption{Plots of envelopes of the residuals $[\da\Ch(0)]^{-1}r(x)$ and
        $[\da\Ch(0)]^{-1}r_c(x)$, each computed from a numerical solution of
        \eqref{eqn:adre_var_loc} for $C(x)$, compared with theoretical
        prediction of the leading-order oscillatory component
        $[\da\Ch(0)]^{-1}\da f(x-k)\Ch(x)$. We have normalised the data in each
        panel by the magnitude of the leading-order term in \eqref{eqn:C_a}.
        There are $N=99$ sinks in all cases, so $\ep = 10^{-2}$; the $(\pe, \da)$
        values are as in Figure~\ref{fig:C_vs_C_H_periodic} and are
        indicated in the panels.  The fine-scale oscillatory structure of residuals is shown in the inset to (b).  The accuracy of the classical [new] method is
    illustrated by the degree of overlap between green [blue] and pink regions (the latter has a dashed boundary).}
    \label{fig:r(x)_periodic}
\end{figure}
We compare the two residuals, $r(x)$ and $\rc(x)$, with the leading-order
theoretical prediction $\da f(x-k)\Ch(x)$ from \eqref{eqn:C_a} in
Figure~\ref{fig:r(x)_periodic} for a range of $(\pe$, $\da)$-values.
All quantities were scaled by $[\da\Ch(0)]^{-1}$ to allow comparison across
parameter values. For illustrative purposes, we plot the envelope of each
function, indicating the fine-scale oscillatory behaviour via $f(x-k)$ in the
inset to Figure~\ref{fig:r(x)_periodic}(b). The present method works
exceptionally well when uptake is weak relative to diffusion and advection
(Figure~\ref{fig:r(x)_periodic}g,h; the blue and pink regions overlap precisely)
and deviates less than the classical method as uptake becomes more significant
(Figure~\ref{fig:r(x)_periodic}a,b,d,e; the blue region overlaps the pink region
better than the green region does). Both methods share the same error when both
advection and diffusion become strong, however
(Figure~\ref{fig:r(x)_periodic}c). Unlike the classical analysis, the present
method does not assume unit-cell periodicity, which is perhaps where this
contribution is lost. As \cite{pavliotisstuart2008} point out in regard to the unit-cell problem,
`the local problem cannot really see the boundary --- this is the key property of scale separation;' 
the present global method avoids this difficulty and is adaptable in principle to parameter ranges 
for which $G$ and $\Ch$ need not be (piecewise) slowly varying on the macroscale.

\FloatBarrier

\section{Stochastic contributions}
\subsection{Normally perturbed sink locations}

We now consider sinks which are weakly perturbed from a periodic arrangement by
normally-distributed random variables so that $\xi_j = j + \sigma \hat\xi_j$,
where $\sigma \ll 1$ and $\hat\xi_j \sim \mathcal{N}(0,1)$. We assume that sinks
do not change places as a result of the random perturbations. Using
\eqref{eqn:C_b_int} and Taylor expanding around the periodic configuration,
\begin{equation}
    \label{eqn:C_b_norm_pert}
    \hat C_b(x) = \da \left\{ \sum_{j \neq k} \left[ \sigma \hat\xi_j\gch_y|_{x,j}
        + \tfrac{1}{2} \sigma^2 \hat\xi_j^2\gch_{yy}|_{x,j} + \dotsb \right] +
    [\gch]_{y=k}^{y=k+\sigma\hat\xi_k} \right\},
\end{equation}
where the cell in which the coordinate $x$ falls, $k = \floor{x+\frac{1}{2}}$,
is again treated separately to avoid expanding non-smooth functions. The
contribution from $\bo{\epm}$ terms in the sum will be an order of magnitude
greater than that from the single unit cell $k$ as $\ep \to 0$, which we
therefore neglect. Assuming that the sinks are independently distributed, which
implies $\Cov(\hat\xi_j,\hat\xi_\ell) = \delta_{j,\ell}$, we write the
covariance as
\begin{equation*}
    \begin{aligned}
        \Cov\bigl(\hat C_b(x_1),\hat C_b(x_2)\bigr) &= \da^2 \sigma^2
        \sum_{j\neq k_1}\sum_{\ell\neq k_2} \gch_y|_{x_1,j}\gch_y|_{x_2,\ell}
        \Cov(\hat\xi_j,\hat\xi_\ell) + \dotsb\\
        &= \da^2 \sigma^2 \sum_{j\notin \{k_1,k_2\}}
        \gch_y|_{x_1,j}\gch_y|_{x_2,j} + \dotsb,
    \end{aligned}
\end{equation*}
where $k_i \equiv \floor{x_i + \frac{1}{2}}$, for $i=1,2$. Using
\eqref{eqn:sum_to_int} to approximate sums with integrals,
\begin{equation}
    \label{eqn:cov_C_b_norm_pert}
    \Cov\bigl(\hat C_b(x_1),\hat C_b(x_2)\bigr) = \da^2 \sigma^2 \int_0^\epm
    \gch_y|_{x_1,y}\gch_y|_{x_2,y} \dd y + \dotsb.
\end{equation}
As observed previously in related problems
\citep{chernyavsky2011transport,chernyavsky2012characterizing,heitzinger2014,russell2016stochastic}
the fluctuations at a given location depend non-locally on the concentration
profile throughout the domain, despite the small and independent perturbations
to the sink locations.

\begin{figure}[t!]
    \centering
    \includegraphics[width=\textwidth]{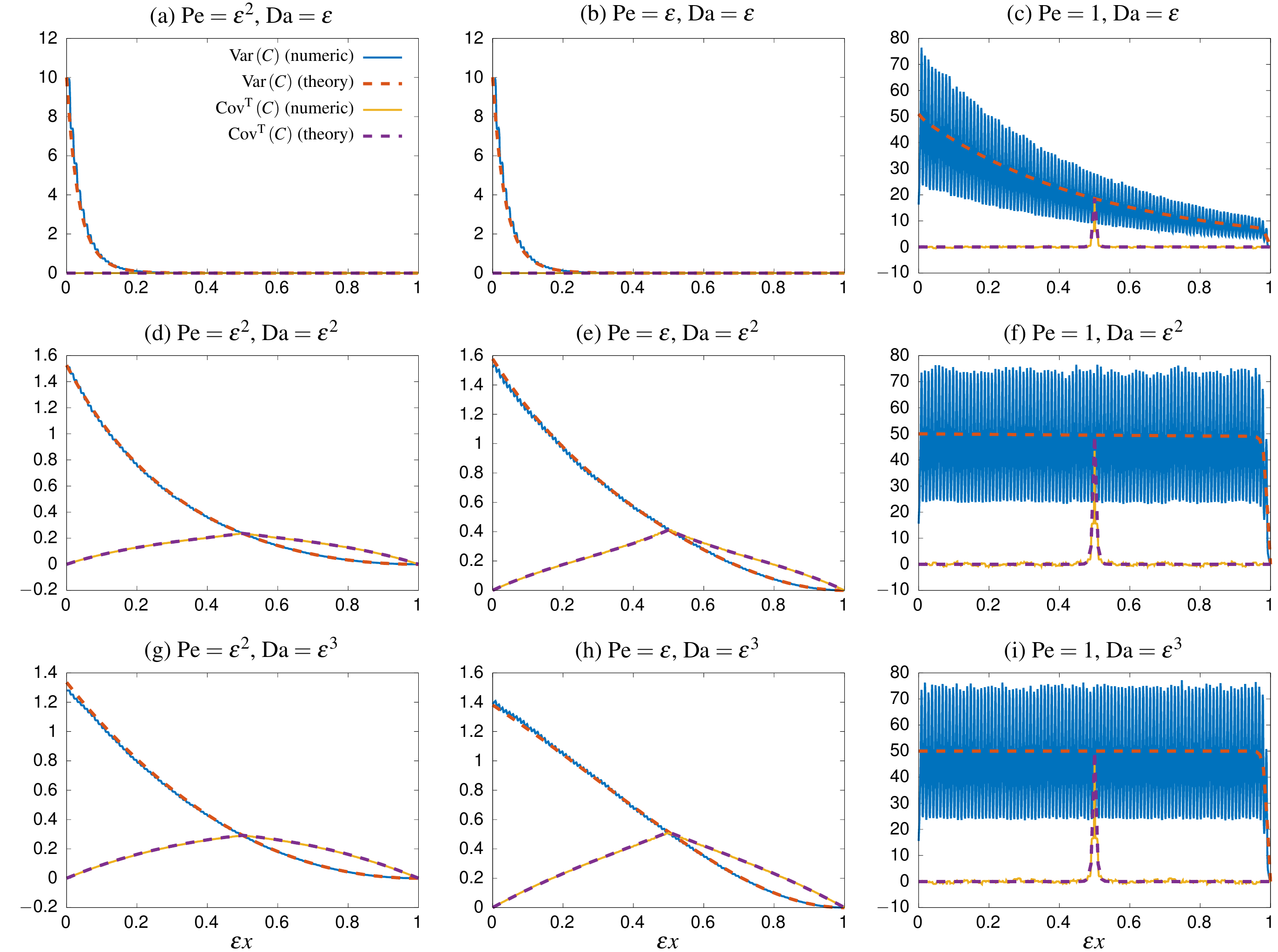}
    \caption{Comparison between the sample variance (solid blue) and transverse
        covariance (solid orange) generated from $10^4$ solutions of
        \eqref{eqn:adre_var_loc} with normally perturbed sink locations, $\xi_j
        = j + \sigma\mathcal{N}(0,1)$, with $\sigma = 0.1$, and the theoretical
        prediction \eqref{eqn:cov_C_b_norm_pert} of variance (dashed red) and
        transverse covariance (dashed purple). All variances and transverse
        covariances have been scaled by $\epm[\da\sigma\Ch(0)G(0,0)]^{-2}$.
        Other parameters are as in Figure~\ref{fig:C_vs_C_H_periodic}.
    }
    \label{fig:var_tcov_norm_pert}
\end{figure}

In Figure~\ref{fig:var_tcov_norm_pert} the theoretical prediction of the
variance, $\var{C(x)} \equiv \Cov\bigl(\hat C_b(x), \hat C_b(x)\bigr)$, and
transverse covariance, $\tcov{C(x)} \equiv \Cov\bigl(\hat C_b(x), \hat
C_b(\epm-x) \bigr)$, given by \eqref{eqn:cov_C_b_norm_pert} are compared with
the sample statistics of an ensemble of $10^4$ Monte-Carlo simulations of the
transport equation \eqref{eqn:adre_var_loc}, computed using the method described
in Appendix~\ref{app:numerics}, with sink locations drawn from a normal
distribution with variance $\sigma^2=10^{-2}$. The agreement is excellent for
$\pe = \bo{\ep}$ or smaller (Figure~\ref{fig:var_tcov_norm_pert}a,b,d,e,g,h).
When advection is stronger, with $\pe = \bo{1}$, sink-to-sink oscillations emerge
in the Monte-Carlo estimates of $\var{C}$. Nevertheless the approximation
\eqref{eqn:cov_C_b_norm_pert} captures its slowly varying mean value, and its
lack of correlation across the domain (reflected by a spike in $\tcov{C}$); we
expect higher-order terms to describe the oscillating part of the variance.

Taking the average of \eqref{eqn:C_b_norm_pert} across realisations of sink
distributions gives
\begin{equation}
    \E{\hat C_b(x)} = \da \left\{\sum_{j\neq k}
        \left[\tfrac{1}{2}\sigma^2\gch_{yy}|_{x,j} + \dotsb\right] +
    \E{[\gch]_{y=k}^{y=k+\sigma\hat\xi_k}} \right\}.
\end{equation}
Again neglecting the small cell-$k$ term and approximating sums with integrals
we find
\begin{equation}
    \label{eqn:mean_C_b_norm_pert}
    \begin{aligned}
        \E{\hat C_b(x)} &= \tfrac{1}{2}\da\sigma^2 \int_0^\epm \gch_{yy}|_{x,y}
        \dd y + \dotsb\\
        &= \tfrac{1}{2}\da\sigma^2 \left[ \gch_y|_{x,\epm} - \gch_y|_{x,x^+} +
        \gch_y|_{x,x^-} - \gch_y|_{x,0} \right] + \dotsb\\
        &= \tfrac{1}{2}\da\sigma^2 \left[ -\Ch(x) - (G {\Ch}_y)|_{x,0} \right] +
        \dotsb,
    \end{aligned}
\end{equation}
where we have applied the jump condition \eqref{eqn:jump_condition} and used
$G_y|_{x,0} = 0$.

\begin{figure}[t!]
    \centering
    \includegraphics[width=\textwidth]{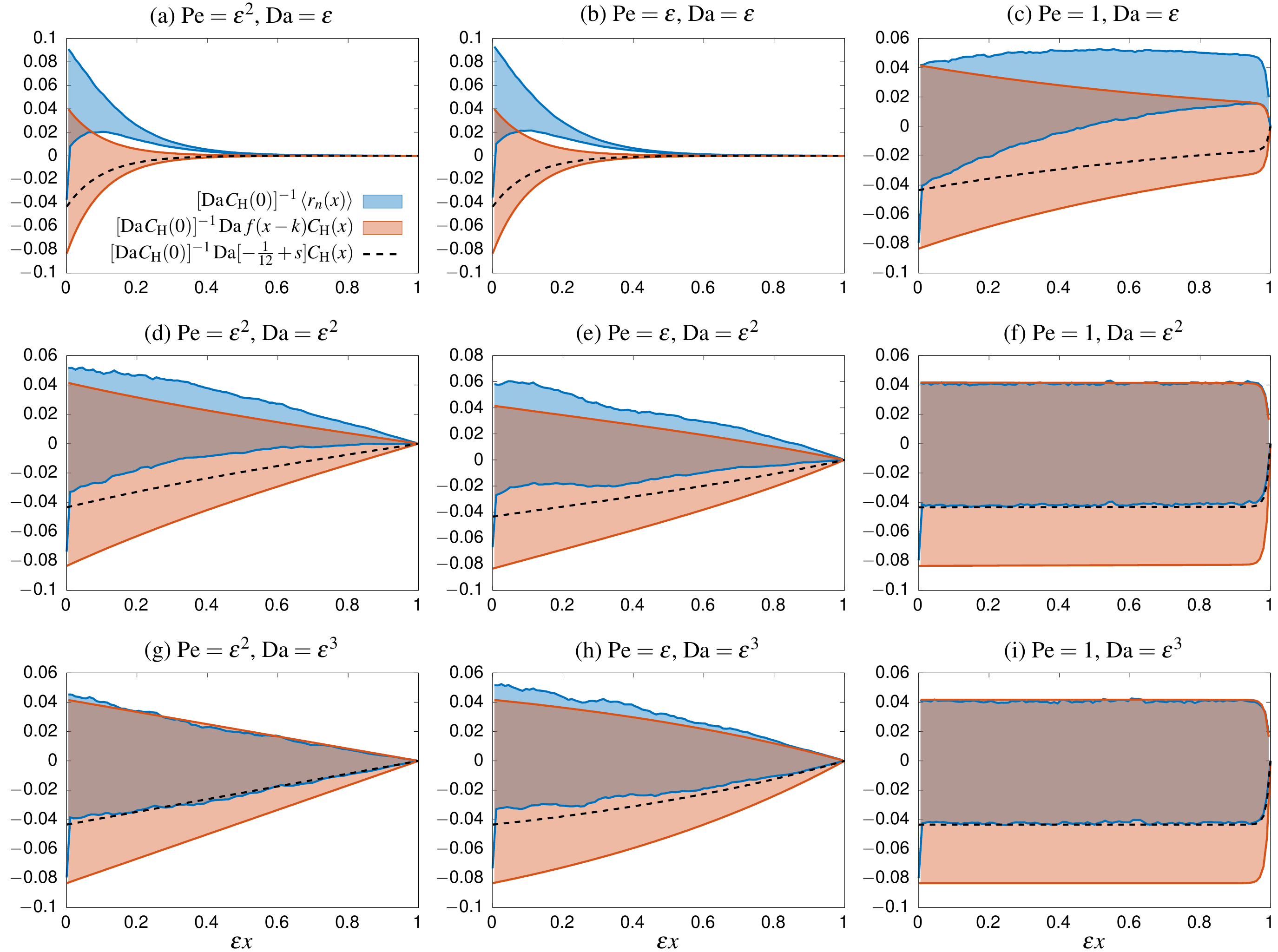}
    \caption{Plots of $[\da\Ch(0)]^{-1}\E{r_n(x)}$, computed from $10^4$
        numerical solutions of \eqref{eqn:adre_var_loc} for $C(x)$ with normally
        perturbed sink locations, $\xi_j = j + \sigma\mathcal{N}(0,1)$, with
        $\sigma = 0.1$, compared with $[\da\Ch(0)]^{-1}\da f(x-k)\Ch(x)$.
        Quantities are normalised in the same way as in
        Figure~\ref{fig:r(x)_periodic}. Other parameters are as in
        Figure~\ref{fig:C_vs_C_H_periodic}. The pink regions match those in
        Figure~\ref{fig:r(x)_periodic}; dashed, black lines show the impact of averaging on their
        lower envelope, derived in
        Appendix~\ref{app:shifted_envelope}, where $s = \sigma/\sqrt{2\pi}$.  The blue regions incorporate the
        correction $\E{\hat{C}_b}$.   The accuracy of the approximation is
        illustrated by the degree of overlap between blue region and the pink region above the dashed line.
    }
    \label{fig:r(x)_norm_pert}
\end{figure}

Using the scaling strategy outlined earlier to analyse (\ref{eqn:cov_C_b_norm_pert},
\ref{eqn:mean_C_b_norm_pert}) in the distinguished limit, we see that the
dominant contribution to the fluctuating part of $\hat C_b$ has magnitude
$\ep^{3/2}\sigma$ and its mean part has magnitude $\ep^2\sigma^2$. We therefore
expect the correction to $\E{C}$ due to stochasticity to be sub-dominant to
$\hat C_a$ provided $\sigma \ll 1$. Figure~\ref{fig:r(x)_norm_pert} compares
the mean residual, 
\begin{equation}
    \E{r_n(x)} = \E{r(x)} + \tfrac{1}{2}\da\sigma^2 \left[ \Ch(x) + (G {\Ch}_y)|_{x,0} \right]
\end{equation} 
(where \eqref{eqn:resid} is modified using \eqref{eqn:mean_C_b_norm_pert}), with
the leading-order oscillatory term, $\da f(x-k) \Ch(x)$. The magnitude and
overall shape of the mean residuals are similar to those from the periodic array
(see Figure~\ref{fig:r(x)_periodic}), although with a slightly reduced amplitude
in some cases. This similarity demonstrates the sub-dominant effect of the weak
stochasticity on the mean compared with the periodic corrections in this
instance. Also shown in Figure~\ref{fig:r(x)_norm_pert}
is the prediction of how the lower envelope of the oscillatory mean residual is elevated as a result of
averaging (see Appendix~\ref{app:shifted_envelope}): averaging an ensemble of `spiky' oscillations of the kind shown in Figure~\ref{fig:r(x)_periodic}(b), with each member of the ensemble displaced laterally by a small normally distributed distance, leads to a smoother mean waveform with an elevated minimum.  This refined lower boundary
agrees very well with simulations for small $\da$, but deviations develop as $\da$ is increased.

Finally, we may use $(\hat{C}_b)_x \vert_{\epm}$ (see (\ref{eq:netflux})) to estimate the uncertainty in the total solute uptake by sinks as 
\begin{equation}
    \Var\bigl((\hat C_b)_x|_{\epm}\bigr) = \da^2 \sigma^2 \int_0^\epm (G_x \Ch)_y|_{\epm,y} \dd y.
\end{equation}

\subsection{Uniformly random sink locations}
We now consider an array of $N$ point sinks, the locations of which are drawn
from a uniform distribution $\mathcal{U}(0,\epm)$, where $\ep = 1/(N+1)$, and
sorted into ascending order: $\xi_1 \le \xi_2 \le \dotsb \le \xi_N$. Thus
$\xi_j$ is the $j$-th order statistic from a set of $N$ uniform random variables
\citep{ahsanullah2013introduction}. In contrast to the array of weakly perturbed
sinks, we cannot Taylor expand around the periodic configuration and neglect the
single unit cell which contains the coordinate $x$. This is because the strong
disorder allows sinks to leave their unit cells and change order. Instead, we
use results on order statistics to approximate the moments of the concentration
profile.

We first note that the sum over all $N$ order statistics $\xi_j$ is equal to the
sum of the $N$ underlying uniform random variables; the former is just a
permutation of the latter. This basic fact is used in \cite{david2003order} and
\cite{chunsheng1992moments} to prove identities involving sums of moments of
order statistics. In particular, we will use the following identity,
\begin{equation}
    \label{eqn:sum_cov_func}
    \mathop{\sum\sum}_{j\neq k}\cov{g(X_{j:N})}{h(X_{k:N})} = \sum_{j=1}^N
    \left[\E{g(X_{j:N})} \E{h(X_{j:N})} - \E{g(X)} \E{h(X)}\right],
\end{equation}
where $X$ is a random variable (with finite variance), $X_{j:N}$ denotes the
$j$-th order statistic out of a sample of size $N$ from the distribution of $X$,
and $g$, $h$ are any real-valued functions satisfying $\var{g(X)} < \infty$,
$\var{h(X)} < \infty$.

Let the uniform random variable $U\sim\mathcal{U}(0,\epm)$ with probability density function $f_U(x) =
\ep$, $0 \le x \le \epm$. Using \eqref{eqn:C_b_int}, we write the covariance as
\begin{equation}
    \begin{aligned}
        \Cov\bigl(\hat C_b(x_1),\hat C_b(x_2)\bigr) &=
        \da^2\sum_{j=1}^N \sum_{k=1}^N
        \cov{\gch_{x_1,\xi_j}}{\gch_{x_2,\xi_k}}\\
        &=\da^2
        \Biggl[
            \mathop{\sum\sum}_{j\neq k}
            \cov{\gch_{x_1,\xi_j}}{\gch_{x_2,\xi_k}} +
            \sum_{j=1}^N\cov{\gch_{x_1,\xi_j}}{\gch_{x_2,\xi_j}}
        \Biggr]\\
        &=\da^2
        \begin{aligned}[t]
            \Biggl\{
                &\sum_{j=1}^N \E{\gch|_{x_1,\xi_j}}\E{\gch|_{x_2,\xi_j}} -
                N\E{\gch|_{x_1,U\vphantom{\xi_j}}}\E{\gch|_{x_2,U\vphantom{\xi_j}}}\\
                +&\sum_{j=1}^N \left[\E{\gch|_{x_1,\xi_j} \gch|_{x_2,\xi_j}} -
                \E{\gch|_{x_1,\xi_j}}\E{\gch|_{x_2,\xi_j}}\right]
            \Biggr\}
        \end{aligned}\\
        &=N\da^2
        \Biggl[
            \E{\gch|_{x_1,U}\gch|_{x_2,U\vphantom{\xi_j}}} -
            \E{\gch|_{x_1,U\vphantom{\xi_j}}}\E{\gch|_{x_2,U\vphantom{\xi_j}}}
        \Biggr],\\
    \end{aligned}
\end{equation}
where we have used \eqref{eqn:sum_cov_func} to transform averages over the
order statistics into averages over the uniform variable $U$.
Writing the averages as integrals using the probability density function $f_U$ then yields
\begin{multline}
    \label{eqn:cov_C_b_uniform}
    \Cov\bigl(\hat C_b(x_1),\hat C_b(x_2)\bigr) =\\
    (1-\ep)\da^2
    \Biggl[
        \int_0^\epm \gch|_{x_1,y} \gch|_{x_2,y} \dd y
        -\ep\int_0^\epm \gch|_{x_1,y} \dd y \int_0^\epm \gch|_{x_2,y} \dd y
    \Biggr].
\end{multline}
We have made no further approximations to arrive at this expression for the
covariance of $\hat C_b$ and it contains contributions at different orders of
$\ep$. A leading-order estimate is obtained by retaining only the $1$ in
$(1-\ep)$, from which we find that the covariance has magnitude $\ep$ in the
distinguished limit, implying fluctuations around the mean have magnitude
$\ep^{1/2}$.

\begin{figure}[t!]
    \centering
    \includegraphics[width=\textwidth]{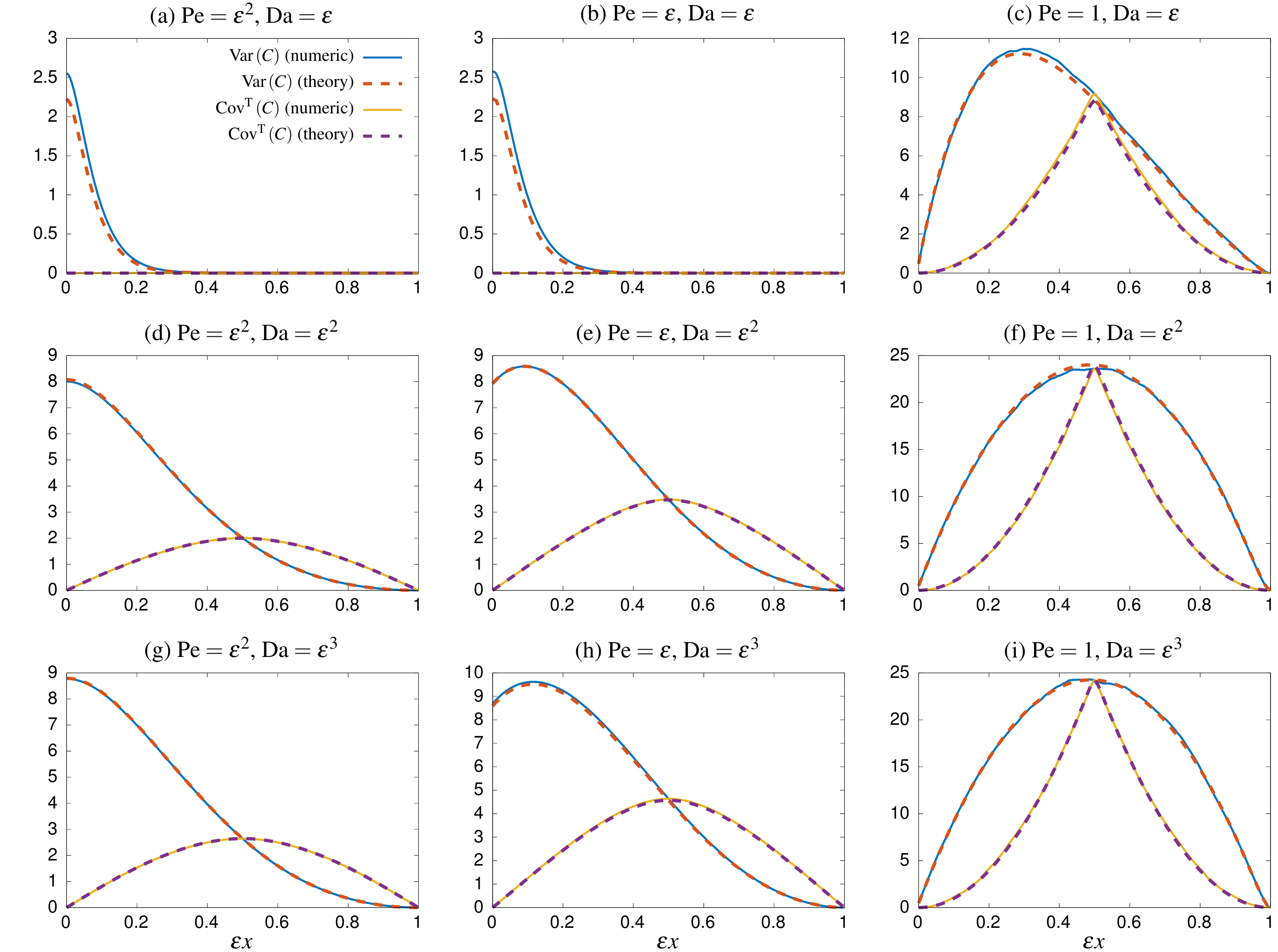}
    \caption{Comparison between the sample variance (solid blue) and transverse
        covariance (solid orange) generated from $10^4$ solutions of
        \eqref{eqn:adre_var_loc} with the $\xi_j$ drawn from
        $\mathcal{U}(0,\epm)$ and then sorted into ascending order for each
        realisation, and the theoretical prediction \eqref{eqn:cov_C_b_uniform}.
        All variances and transverse covariances have been scaled by
        $[\da\Ch(0)G(0,0)]^{-2}$. Other parameters are as in
        Figure~\ref{fig:C_vs_C_H_periodic}.
    }
    \label{fig:var_tcov_uniform}
\end{figure}

In Figure~\ref{fig:var_tcov_uniform} we compare the theoretical prediction
of the variance and transverse covariance, in \eqref{eqn:cov_C_b_uniform}, with
sample statistics of an ensemble of $10^4$ Monte-Carlo simulations of
\eqref{eqn:adre_var_loc}, with sink locations drawn from a uniform distribution
$\mathcal{U}(0,\epm)$ and labelled in ascending order. For all values of $(\pe,
\da)$ shown, the agreement is excellent, except perhaps for a small discrepancy
near the inlet when $\da = \bo{\ep}$ and $\pe \ll 1$. In contrast to
Figure~\ref{fig:var_tcov_norm_pert}, sink-to-sink oscillations in the variance
or transverse covariance are not visible here. We expect that oscillations
appear at higher order and are sub-dominant to the effects of strongly
disordered sink locations.

The variance in the net uptake by sinks, from (\ref{eq:netflux}), is derived analogously to (\ref{eqn:cov_C_b_uniform}) as
\begin{equation}
    \Var\bigl((\hat C_b)_x|_\epm \bigr) =\da^2
    \Biggl[
        \int_0^\epm \left[ (G_x \Ch)|_{\epm,y} \right]^2 \dd y
        - \ep\left[\int_0^\epm (G_x \Ch)|_{\epm,y} \dd y\right]^2 
    \Biggr].
\end{equation}
Interestingly, simulations show that $(\hat C_b)_x \vert_\epm$ has a roughly Gaussian distribution for $\pe=\bo{\ep}$ and $\da=\bo{\ep^2}$, but an asymmetric distribution when $\pe=\bo{1}$ (not shown).

We now turn to the mean of the first stochastic correction. Using again the
equivalence of sums over order statistics and their underlying random variables,
the mean of $\hat C_b$ can be written as
\begin{equation}
    \begin{aligned}
        \E{\hat C_b(x)} &= \da \sum_{j=1}^N \left[ \E{\gch_{x,\xi_j}} - \gch|_{x,j}
        \right]\\
        &=\da \sum_{j=1}^N
        \left[ \E{\gch_{x,U\vphantom{\xi_j}}} - \gch|_{x,j} \right]\\
        &=\da 
        \left[(1-\ep)\int_0^\epm \gch|_{x,y}\dd y - \sum_{j=1}^N \gch|_{x,j}\right].
    \end{aligned}
\end{equation}
In the distinguished limit, the first integral in the brackets is
$\bo{\ep^{-2}}$, the second integral is $\bo{\epm}$, and the sum is
$\bo{\ep^{-2}}$. Converting the sum to an integral, we expect the
$\bo{\ep^{-2}}$ contributions to cancel the first integral, so we retain
terms up to the size of the smaller integral term,
\begin{equation}
    \begin{aligned}
        \sum_{j=1}^N \gch|_{x,j} &= \int_0^\epm \gch|_{x,y} \dd y - \tfrac{1}{2}
        \left[ \gch|_{x,0} + \gch|_{x,\epm} \right] + \dotsb\\
        &= \int_0^\epm \gch|_{x,y} \dd y + \tfrac{1}{2} \epm\Ch(0)\Ch(x) +
        \dotsb,
    \end{aligned}
\end{equation}
where we have applied the outlet boundary condition \eqref{eqn:adre_var_loc_bcs}
and \eqref{eqn:C_H_G+}. Using this approximation, we find
\begin{equation}
    \label{eqn:C_b_mean_uniform_smooth}
    \E{\hat C_b(x)} = \da\left[ -\ep\int_0^\epm \gch|_{x,y} \dd y -
    \tfrac{1}{2}\epm\Ch(0)\Ch(x) + \bo{1} \right],
\end{equation}
which has magnitude $\bo{\ep}$ in the distinguished limit.

\begin{figure}[t!]
    \centering
    \includegraphics[width=\textwidth]{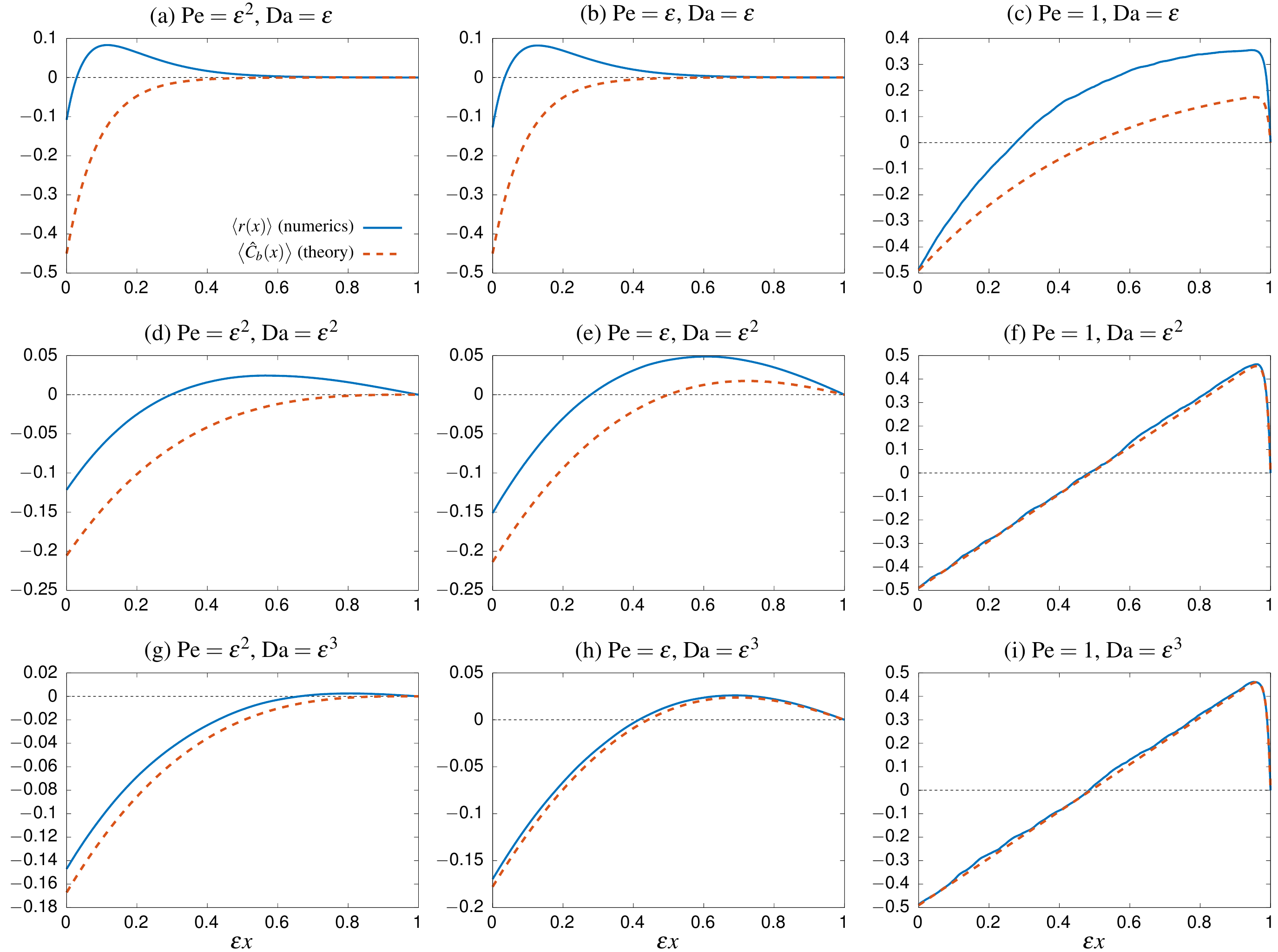}
    \caption{Plots of $\E{r(x)}$, each computed from $10^4$ numerical solutions
        of \eqref{eqn:adre_var_loc} for $C(x)$ with uniformly distributed sinks,
        $\xi_j$ drawn from $\mathcal{U}(0,\epm)$, and the theoretical prediction
        \eqref{eqn:C_b_mean_uniform_smooth}. All quantities have been scaled by
        $\ep\da^{-1}\Ch(0)^{-2}$. Other parameters are as in
        Figure~\ref{fig:C_vs_C_H_periodic}. 
    }
    \label{fig:r(x)_uniform}
\end{figure}

Figure~\ref{fig:r(x)_uniform} compares the mean residual $\langle r(x)\rangle$,
calculated using \eqref{eqn:resid} from $10^5$ Monte-Carlo samples, each with
sink locations drawn from a uniform distribution, and the theoretical prediction
\eqref{eqn:C_b_mean_uniform_smooth}. We compare the sample statistics with this
$\bo{\ep}$ stochastic contribution rather than the $\bo{\ep^2}$ oscillating
part, $\da f(x-k)\Ch(x)$, since the sink-to-sink oscillations appear at higher
order here. The theory predicts the overall magnitude and shape of the
correction to the mean concentration for many parameter values, especially when
$\da$ is small, and when $\pe$ is large. However, some features, such as near
the inlet boundary in Figure~\ref{fig:r(x)_uniform}(a,b) are not captured by the
leading-order theory for this choice of $\ep$; we expect that further correction terms will account for
these discrepancies.

\subsection{Magnitude estimates in other regions of parameter space}

Following \cite{russell2016stochastic}, we can identify three parameter regimes
around the distinguished limit. In each regime, either diffusion [D], advection
[A] or uptake [U] is the dominant process. These can be identified by balancing
the various terms in \eqref{eqn:C_H_eqn_bcs}. [D] is defined by the region $\pe
\ll \ep$, $\da \ll \ep^2$; [A] by $\pe \gg \ep$, $\da \ll \pe^2$; and [U] by
$\da \gg \max(\ep^2,\pe^2)$. In [D], $\Ch$ varies smoothly across the whole
domain over a lengthscale $x\sim \varepsilon^{-1}$ (see
Figure~\ref{fig:C_vs_C_H_periodic}(g)). In [U], the lengthscale shrinks to
$x\sim \da^{-1/2}$ (Figure~\ref{fig:C_vs_C_H_periodic}a,b). In [A], for $
\varepsilon \pe\ll \da \ll \pe^2$ this lengthscale increases to $x\sim \pe/\da$ (we
denote this subregion [A$_I$]), and then encompasses the whole domain for $\da
\ll \varepsilon \pe$ (with a short boundary layer of length $x\sim 1/\pe$ at the
outlet; this is subregion [A$_{II}$], see
Figure~\ref{fig:C_vs_C_H_periodic}(f,i)). Other panels in
Figure~\ref{fig:C_vs_C_H_periodic} sit at interfaces between these regions: (c)
[A$_I$/A$_{II}$]; (d) [D/U]; (e) [A/D/U], the central distinguished limit; and
(h) [D/A]. We use these lengthscales to estimate the asymptotic magnitudes of
$\Ch$ and $G$ in these regions of parameter space, from which we deduce the
magnitudes of the deterministic correction $\hat C_a$, $\hat C_{aa}$ in terms of
$\Ch$ and $G$. These results are summarised in
Table~\ref{tab:mag_C_h_G_C_a_C_aa}. 

The macroscale approximation is slowly varying when $\pe\ll 1$
(for the downstream boundary layer to extend over multiple sinks) and $\da \ll
1$ (ensuring the corrections due to discreteness to remain subdominant to
$\Ch$). Oscillatory corrections in $\hat{C}_a$ grow as each of these boundaries in parameter space
is approached, indicating how $\Ch$ becomes an increasingly poor approximation
of the gradient of the exact solution over short lengthscales.

\begin{table}[t]
    \centering
    \begin{tabular}{c|c c c c}
        Region & $\Ch$ & G & $\hat C_a$ & $\hat C_{aa}$\\ \hline
        A/D/U & 1 & $\epm$ & $\ep(1,\ep)$ & $\ep^2$\\
        D & 1 & $\epm$ & $\da\epm(1,\ep)$ & $\da^2\ep^{-2}$\\
        U & $\ep\da^{-1/2}$ & $\da^{-1/2}$ & $\ep(1,\da^{1/2})$ &
        $\ep\da^{1/2}$\\
        A & $\ep\pe^{-1}$ & $\pe^{-1}$ & $\ep\da\pe^{-2}(1,\pe)$ &
        $\ep\da^2\pe^{-3}$
    \end{tabular}
    \caption{Asymptotic magnitudes of the leading-order homogenized solution
        $\Ch$ and the Green's function $G$ in various parameter regimes (defined
        in the text), with corresponding estimates of the magnitudes of the
        deterministic corrections $\hat C_a$ and $\hat C_{aa}$. In the $\hat
        C_a$ column, the first term in the parentheses corresponds to the
        leading-order slowly varying terms and the second to the amplitude of
    leading-order oscillatory terms.}
    \label{tab:mag_C_h_G_C_a_C_aa}
\end{table}

Turning to the stochastic corrections, we estimate the asymptotic magnitudes of
the mean and fluctuations (given by the standard deviation) of $\hat C_b$ for
both normally perturbed and uniformly-random sink locations. These magnitudes
are summarised in Table~\ref{tab:mag_C_b}. It turns out that the domain of
validity of $\Ch$ remains $\da \ll 1$ and $\pe \ll 1$ in all cases. However in
the uniformly-random case, the dominant correction to $\Ch$ is due to
randomness, whereas for normally perturbed sinks, the discrete correction
dominates the stochastic one.

\begin{table}[t]
    \centering
    \begin{tabular}{c|c c|c c}
        \multicolumn{1}{c|}{} & \multicolumn{2}{c|}{Normally perturbed} &
        \multicolumn{2}{c}{Uniform}\\[1em]
        Region & $\E{\hat C_b}$ & $\sqrt{\var{\hat C_{b}}}$ & $\E{\hat C_b}$ & $\sqrt{\var{\hat C_{b}}}$\\\hline
        A/D/U & $\sigma^2\ep^2$ & $ \sigma\ep^{3/2}$ & $\ep$ & $\ep^{1/2}$\\
        D & $\sigma^2\da$ & $\sigma\ep^{-1/2}\da$ & $\epm\da$ & $\ep^{-3/2}\da$\\
        U & $\sigma^2\ep\da^{1/2}$ & $\sigma\ep \da^{1/4}$
        & $\ep$ & $\ep\da^{-1/4}$
        \\
        A$_{I}$ & $\sigma^2\ep\da\pe^{-1}$ & $\sigma\ep^{3/2}\da\pe^{-2}$ & $\ep\da\pe^{-2}$ & $\ep^{1/2}\da\pe^{-2}$ \\
        A$_{II}$ & $\sigma^2\ep\da\pe^{-1}$ & $\sigma\ep \da^{3/2}\pe^{-5/2}$ & $\ep\da\pe^{-2}$ & $\ep \da^{1/2}\pe^{-3/2}$
            \end{tabular}
    \caption{Magnitude estimates of the mean and fluctuations of the stochastic
    correction $\hat C_b$}
    \label{tab:mag_C_b}
\end{table}

\section{Discussion}
We have analysed a model of transport past an array of point sinks with
first-order uptake kinetics. We considered periodic, weakly perturbed and
strongly disordered arrays; weak disorder was modelled using small normally-distributed perturbations from a periodic configuration while strongly
disordered sinks had uniformly-randomly distributed locations. We posed an
\textit{ad hoc} expansion for the concentration field, centred around the
leading-order homogenized concentration (which is obtained using classical two-scale
methods described in Appendix~\ref{app:periodic_homog}), in which the
higher-order terms can be identified as corrections due to the discrete nature
of the sinks and the effects of disorder, and combinations thereof. However, the
asymptotic ordering of the expansion is not known \textit{a priori} and it
contains a mixture of expressions varying on long and short lengthscales and
having deterministic and stochastic components. We assessed the magnitudes of
the different correction terms in the distinguished limit $\pe = \bo{\ep}$, $\da
= \bo{\ep^2}$ for each sink distribution. This process elucidates whether the
dominant corrections to the homogenized mean concentration profile arise from
discreteness of the sinks (as is the case for normally perturbed sinks with
$\sigma \ll 1$) or from spatial disorder (in the uniformly-random case).
Likewise, our results demonstrate when fluctuations in the concentration become
comparable in size to the mean, signifying a breakdown of the expansion.
Interestingly, for the present problem the homogenized approximation holds for
$\da \ll 1$ (with $\pe \ll 1$), breaking down in region [U] as $\da$ approaches
unity both in the strictly periodic case (when the dominant relative error is
$\bo{\da^{1/2}}$, see Table~\ref{tab:mag_C_h_G_C_a_C_aa}) and the uniformly
random case (when the dominant relative error is $\bo{\da^{1/4}}$, see
Table~\ref{tab:mag_C_b}).

Our results provide evidence that for a periodic sink array, the classical
method employing a two-scale expansion and a unit-cell average (summarized in
Appendix~\ref{app:periodic_homog}) fails to account accurately for higher-order
corrections in the concentration field. The alternative method presented here
neither assumes that the concentration explicitly depends on two spatial
variables nor that it is periodic across unit cells, and it recovers terms
missing in the classical approach that improve agreement with numerical
simulations (up to a given order of $\ep$), as illustrated in
Figure~\ref{fig:r(x)_periodic}. Limitations of the two-scale method in dealing with boundary conditions have been noted previously by \cite{pavliotisstuart2008}.

When sink locations are weakly perturbed, the concentration field has a spatial
correlation structure that extends across the entire domain, even though the
perturbations to the sink locations are independent. Our Green's function-based
approach provides an explicit prediction of these correlations in terms of a
non-local combination of $G$ and the leading-order concentration profile $\Ch$
which agrees well with simulations in a large region of $(\pe, \da)$-parameter
space (Figure~\ref{fig:var_tcov_norm_pert}). In the present problem the first
corrections to the mean concentration that result from weakly disordered sinks
are smaller in magnitude than the corrections due to discreteness (arising in
the periodic problem), provided $\sigma \ll 1$. This is confirmed by
comparison with simulations of the ensemble averaged residual
$\E{r(x)}$ (see Figure~\ref{fig:r(x)_norm_pert}).
In our previous study of the case in which sink strength, rather than sink
location, was disordered, we found that fluctuation magnitudes could be greatest
towards the downstream end of the domain when advection was strong
\citep{russell2016stochastic}, unlike the pattern of disorder shown in
Figure~\ref{fig:var_tcov_norm_pert}.

Strongly disordered sink locations were modelled using a uniform distribution
$\mathcal{U}(0,\epm)$ and labelled in ascending order from the inlet to the
outlet of the domain. The location of the $j$-th sink, $\xi_j$, is therefore the
$j$-th order statistic of the uniform distribution out of $N$. Before
relabelling, the locations are independent random variables but the sorting
introduces correlations between the sink locations. Using results on order
statistics, we derived predictions of the long-range correlations in the
concentration induced by sink disorder, which again agree well with simulations
(see Figure~\ref{fig:var_tcov_uniform}). Unlike weakly perturbed sinks, we found
that strong disorder has a significant effect on the mean concentration,
contributing at $\bo{\ep}$ in the distinguished limit (Table~\ref{tab:mag_C_b}). This is an order of
magnitude larger than the oscillatory terms arising from a periodic array, and
we therefore compare the mean stochastic corrections with the sample mean of the
residual $\E{r(x)}$ from simulations, rather than with $\da f(x-k) \Ch(x)$ as
before (see Figure~\ref{fig:r(x)_uniform}). Our estimate of the perturbation to
the mean concentration induced by disorder shows excellent quantitative accuracy
for smaller values of $\da$; the relative error in the mean is $\bo{\da^{1/2}}$
as uptake becomes stronger, and this grows as $\da$ increases towards unity.
Unlike the case of variable sink strength \citep{russell2016stochastic}, here
the correction to the mean due to disordered sink locations can change sign across the
domain.

In practical applications it can be important to understand not only large-scale
concentration distributions across a region but also small-scale variations
across unit cells. In the placenta, for example, transfer between fetal and
maternal circulation takes place at the lengthscale of individual terminal villi, 
where individual fetal capillary loops within a branch come into close
proximity to maternal blood outside the branch \citep{erlich2018}. The size of solute fluctuations
across an individual branch can be expected to influence the transport across
the surface of the branch. Given the high degree of spatial disorder in
branches \citep{chernyavsky2011transport, erlich2018}, the fluctuations associated with
spatial disorder (reflected by the standard deviation of $\hat{C}_b$ in
Table~\ref{tab:mag_C_b}) deserve particular attention, particularly if there are
correlations between the orientation of capillary loops within the villous
branch and the position of the branch with respect to its neighbours.

There are a number of obvious extensions of the present work, for example to
consider other distinguished limits in parameter space, nonlinear uptake kinetics and unsteady effects.
A similar steady problem with zeroth-order kinetics was analysed in
\cite{chernyavsky2011transport,chernyavsky2012characterizing} using a direct
algebraic method to capture stochastic behaviour. A Green's function approach
may be applicable to such cases but the effort in calculating nonlinear, nonlocal and unsteady
expressions is likely to be greater than in the present case.
A further important class of problems to consider involves finite-size sinks in
two or more dimensions, where there has been substantial effort in deriving
upscaled approximations for electrokinetics \citep{heitzinger2014, schmuck2015} and reactive flow in disordered porous media
\citep{Cushman2002}. The present approach should provide a useful foundation for
investigations characterising the multiscale structure of stochastic flow and
solute fields in higher dimensions.

\section*{Acknowledgements}

We are grateful to Tobias Galla for helpful conversations.  OEJ was supported by EPSRC grant EP/K037145/1.

\appendix

\section{Numerics}
\label{app:numerics}
In this section we describe a hybrid method for generating realisations of the
concentration profile for arbitrary sink distributions. We represent solutions
of \eqref{eqn:adre_var_loc} exactly as an algebraic linear system, which we
solve numerically for a given sink distribution.

We first write \eqref{eqn:adre_var_loc_eqn} as an advection-diffusion equation
between sinks and a condition in the jump in concentration gradient at each
sink. Integrating in a vanishing region around sink $j$ and using continuity of
the concentration, gives the jump condition
\begin{equation}
    \label{eqn:jump_condition}
    \jump{C_x}_{x=\xi_j} = \da C(\xi_j), \quad j = 1, \dotsc, N,
\end{equation}
whereas in the bulk of the domain,
\begin{equation}
    \label{eqn:bulk_equation}
    C_{xx} - \pe C_x = 0, \quad 0 < x < \epm, \quad x \neq \xi_j,
\end{equation}
supplemented with the boundary conditions \eqref{eqn:adre_var_loc_bcs} and
continuity of $C$ across sinks, $\jump{C}_{x=\xi_j} = 0$.
Integrating \eqref{eqn:bulk_equation} twice, we find
\begin{equation}
    \label{eqn:C_bulk}
    C(x) = A_j e^{\pe(x-\xi_j)} + B_j, \quad \xi_j \le x \le \xi_{j+1}, \quad
    j = 0, \dotsc, N,
\end{equation}
where the $A_j$, $B_j$ are constants. The inlet and outlet boundary conditions
yield
\begin{equation*}
    B_0 = \ep/\pe, \quad A_N e^{\pe(\epm-\xi_N)} + B_N = 0,
\end{equation*}
respectively.
\begin{subequations}
    \label{eqn:A_j_B_j_lin_sys_pre}
    Continuity of $C$ across sinks implies,
    \begin{equation}
        A_j - A_{j-1} e^{\pe(\xi_j-\xi_{j-1})} + B_j - B_{j-1} = 0,
    \end{equation}
    and the jump condition \eqref{eqn:jump_condition} gives,
    \begin{equation}
        A_j - A_{j-1} e^{\pe(\xi_j-\xi_{j-1})} - \frac{\da}{\pe}\left(A_j + B_j\right) = 0.
    \end{equation}
\end{subequations}
Eliminating $A_{j-1}$ from (\ref{eqn:A_j_B_j_lin_sys_pre}b) via
(\ref{eqn:A_j_B_j_lin_sys_pre}c), gives the following sparse linear system of
$2(N+1)$ algebraic equations for the $A_j$, $B_j$,
\begin{subequations}
    \label{eqn:A_j_B_j_lin_sys}
    \begin{align}
        B_0 &= \frac{\ep}{\pe},\\
        \frac{\da}{\pe}A_j + \left(1 + \frac{\da}{\pe}\right)B_j - B_{j-1} &=
        0, \qquad j=1,\dotsc,N\\
        \left(1-\frac{\da}{\pe}\right)A_j - e^{\pe(\xi_j-\xi_{j-1})}A_{j-1} -
        \frac{\da}{\pe}B_j &= 0, \qquad j=1,\dotsc,N\\
        A_N e^{\pe(\epm-\xi_N)} + B_N &= 0.
    \end{align}
\end{subequations}
This is an exact representation of \eqref{eqn:adre_var_loc} but an explicit
solution for the $A_j$, $B_j$ is not readily available. Therefore, for a given
sink distribution, we find $A_j$, $B_j$ via a numerical solution of the linear
system \eqref{eqn:A_j_B_j_lin_sys}, and use the coefficients in the bulk
solution \eqref{eqn:C_bulk} to reconstruct the concentration profile. An
ensemble of such concentration profiles with sinks placed according to some
random distribution can be used to generate sample statistics which we will use
to validate theoretical predictions.

\section{Classical two-scale homogenization for transport past a periodic sink array}
\label{app:periodic_homog}
We use a classical homogenization approach to analyse the
periodic sinks problem for comparison with the method described in Sec.~\ref{sec:periodic}. We begin with \eqref{eqn:adre_var_loc} with $\xi_j=j$, \ie
\begin{subequations}
    \label{eqn:adre}
    \begin{gather}
        C_{xx} - \pe \, C_x = \da \, C(x) \, S(x), \quad 0 < x < \epm,\\
        \pe \, C|_{x=0} - C_x|_{x=0} = \ep,\\
        C|_{x=\epm} = 0,\\
        S(x) = \sum_{i=1}^N \delta(x-i)\label{eqn:adre_sink_dist}.
    \end{gather}
\end{subequations}
Let $X = \ep x$ denote a ``long-range'' spatial variable which takes values
in $[0,1]$ across the domain. We then seek solutions of the form $C(x) = \ti
C(x,\ep x)$, where we expand in powers of $\ep$,
\begin{equation}
    \label{eqn:Ctilde_exp}
    \ti C(x,X) = \C0(x,X) + \ep\C1(x,X) + \ep^2\C2(x,X) + \dotsb,
\end{equation}
where $\C n(x,X) = \bo 1$, for $n=0,1,2,\dotsc$. The variables $x$ and $X$
will be treated as independent. Derivatives transform according to
\begin{equation}
    \diff{}{x} = \pdiff{}{x} + \ep \pdiff{}{X}, \qquad \diff{^2}{x^2} =
    \pdiff{^2}{x^2} + 2\ep\pdiff{^2}{x\partial X} + \ep^2\pdiff{^2}{X^2}.
\end{equation}

We investigate solutions in the distinguished limit $\pe = \bo\ep$, $\da =
\bo{\ep^2}$. Thus, we set $\pe = \ep p_0$ and $\da = \ep^2 q_0$, where $p_0,
q_0 = \bo 1$ as $\ep \to 0$. Define the linear operator and boundary
condition operator as
\begin{equation}
    \lapp \equiv \pdiff{^2}{x^2},\qquad
    \bapp C \equiv \{C_x|_{x=0}, C|_{X=1}\},
\end{equation}
respectively. We have included the subscripts to distinguish these operators
from those in the main text. Then substituting \eqref{eqn:Ctilde_exp} into
\eqref{eqn:adre} and collecting terms in orders of $\ep$, we obtain the
following systems:
\begin{subequations}
    \label{eqn:C_problem_all}
    \begin{align}
        &\begin{gathered}
            \label{eqn:C_problem_O(1)}
            \bo 1:\quad
            \left\{
                \begin{aligned}
                    \lapp\C0 &= 0\\
                    \bapp{\C0} &= \{0,0\},
                \end{aligned}
            \right.
        \end{gathered}\\[1em]
        &\begin{gathered}
            \label{eqn:C_problem_O(ep)}
            \bo\ep:\quad
            \left\{
                \begin{aligned}
                    \lapp\C1 &= -2\C0_{xX} + p_0 \C0_x\\
                    \bapp{\C1} &= \{p_0 \C0\bigr|_{X=0} - \C0_X\bigr|_{X=0} - 1, 0\},
                \end{aligned}
            \right.
        \end{gathered}\\[1em]
        &\begin{gathered}
            \label{eqn:C_problem_O(ep2)}
            \bo{\ep^2}:\quad
            \left\{
                \begin{aligned}
                    \lapp\C2 &= q_0 \C0 S(x) - 2\C1_{xX} - \C0_{XX} + p_0
                    \left(\C1_x + \C0_X\right)\\
                    \bapp{\C2} &= \{p_0 \C1\bigr|_{X=0} - \C1_X\bigr|_{X=0}, 0\},
                \end{aligned}
            \right.
        \end{gathered}\\[1em]
        &\begin{gathered}
            \label{eqn:C_problem_O(ep3)}
            \bo{\ep^3}:\quad
            \left\{
                \begin{aligned}
                    \lapp\C3 &= q_0 \C1 S(x) - 2\C2_{xX} - \C1_{XX} + p_0
                    \left(\C2_x + \C1_X\right)\\
                    \bapp{\C3} &= \{p_0 \C2\bigr|_{X=0} - \C2_X\bigr|_{X=0}, 0\},
                \end{aligned}
            \right.
        \end{gathered}
    \end{align}
\end{subequations}
and so on for higher orders.

We seek $x$-periodic solutions and therefore we will work in a
representative unit cell by defining new coordinates, $x' \equiv x - i$,
so that $-1/2 < x' < 1/2$ in each cell. Sinks are therefore situated at $x'
= 0$ in each cell.

At $\bo 1$, we have
\begin{subequations}
    \begin{gather}
        \C0_{x'x'} = 0,\\
        \C0_{x'}\bigr|_{X=0} = 0, \quad \C0\bigr|_{X=1} = 0,\\
        \jump{\C0}_{x'=0} = 0, \quad \jump{\C0_{x'}}_{x'=0} = 0,\\
        \text{$x'$-periodicity of } \C0.
    \end{gather}
\end{subequations}
Using the periodicity condition, we therefore find that $\C0 = \C0(X)$ only,
along with $\C0(1) = 0$.

At $\bo\ep$, we have
\begin{subequations}
    \begin{gather}
        \C1_{x'x'} = 0,\\
        \C1_{x'}\bigr|_{X=0} = \C0_B, \quad \C1\bigr|_{X=1} = 0,\\
        \jump{\C1}_{x'=0} = 0, \quad \jump{\C1_{x'}}_{x'=0} = 0,\\
        \text{$x'$-periodicity of } \C1,
    \end{gather}
\end{subequations}
where $\C0_B \equiv p_0 \C0\bigr|_{X=0} - \C0_X\bigr|_{X=0} - 1$. Similarly
to the previous order, we find that $\C1 = \C1(X)$ only, with $\C1(1) = 0$.
Additionally, applying the boundary condition on $\C1$ at the inlet fixes the
condition on $\C0$ to be $\C0_B = 0$, or
\begin{equation}
    \label{eqn:C0_inlet_bc}
    p_0 \C0\bigr|_{X=0} - \C0_X\bigr|_{X=0} = 1.
\end{equation}

At $\bo{\ep^2}$, we have
\begin{subequations}
    \begin{gather}
        \C2_{x'x'} = -\left(\C0_{XX} - p_0 \C0_X\right),\\
        \C2_{x'}\bigr|_{X=0} = \C1_B, \quad \C2\bigr|_{X=1} = 0,\\
        \jump{\C2}_{x'=0} = 0, \quad \jump{\C2_{x'}}_{x'=0} = q_0 \C0,\\
        \text{$x'$-periodicity of } \C2,
    \end{gather}
\end{subequations}
where $\C1_B \equiv p_0 \C1\bigr|_{X=0} - \C1_X\bigr|_{X=0}$. Performing a
spatial average over a unit cell and using the periodicity and jump conditions
on $\C2$, we obtain the macroscopic equation for the leading order solution,
\begin{equation}
    \label{eqn:C0_eqn}
    \C0_{XX} - p_0 \C0_X = q_0 \C0(X).
\end{equation}
The leading order solution is
\begin{equation}
    \label{eqn:C0_solution}
    \C0(X) = \frac{e^{p_0 X/2} \sinh[\Phi(1-X)]}{p_0\sinh(\Phi)/2 + \Phi
    \cosh(\Phi)},
\end{equation}
where $\Phi \equiv \sqrt{q_0 + p_0^2/4}$, consistent with \eqref{eqn:C_H}.

Substituting the macroscopic equation for $\C0$ into the equation for $\C2$ and
directly integrating in each half of the unit cell yields
\begin{equation}
    \C2 = \left\{
    \begin{alignedat}{2}
        &-\tfrac{1}{2}q_0 \C0 {x'}^2 + a_1 x' + a_2, & -\tfrac{1}{2} \le {}&x' < 0,\\
        &-\tfrac{1}{2}q_0 \C0 {x'}^2 + b_1 x' + b_2, \qquad & 0 \le {}&x' \le
        \tfrac{1}{2}.
    \end{alignedat}
    \right.
\end{equation}
Applying the jump conditions gives $b_2 = a_2$ and $b_1 = a_1 + q_0 \C0$. We
then impose that $\E{\C2}$ is a constant, where $\E{ f } = \int_{-1/2}^{1/2} f
\dd x'$, which gives $a_2 = \E{\C2} -\frac{1}{12}q_0 \C0$. Periodicity then
implies that $a_1 = -\frac{1}{2} q_0 \C0$.

It remains to find $\C1_B$ and
$\E{\C2}$ which allow $\C2$ to satisfy the global boundary conditions.
We therefore find from the inlet condition that
\begin{equation}
    \C2_{x'}\bigr|_{X=0} = -\tfrac{1}{2} q_0 \C0\bigr|_{X=0}(-1) = \C1_B,
\end{equation}
or
\begin{equation}
    \C1_B = \tfrac{1}{2} q_0 \C0(0).
\end{equation}
Similarly, applying the outlet condition gives $\E{\C2} = 0$. Therefore the
expression for the second correction is
\begin{equation}
    \C2 =-\tfrac{1}{2}q_0 \C0\left({x'}^2 - |x'| + \tfrac{1}{6}\right), \quad
    -\tfrac{1}{2} \le x' \le \tfrac{1}{2}.
\end{equation}

At $\bo{\ep^3}$, we have
\begin{subequations}
    \begin{gather}
        \C3_{x'x'} = -\left( 2\C2_{x'X} - p_0 \C2_{x'} \right) -\left(\C1_{XX} - p_0 \C1_X\right),\\
        \C3_{x'}\bigr|_{X=0} = \C2_B, \quad \C3\bigr|_{X=1} = 0,\\
        \jump{\C3}_{x'=0} = 0, \quad \jump{\C3_{x'}}_{x'=0} = q_0 \C1,\\
        \text{$x'$-periodicity of } \C3,
    \end{gather}
\end{subequations}
where $\C2_B \equiv p_0 \C2\bigr|_{X=0} - \C2_X\bigr|_{X=0}$. Again we
average over a unit cell, and find that the first correction satisfies the same
macroscopic equation as the leading order solution,
\begin{equation}
    \C1_{XX} - p_0 \C1_X = q_0 \C1(X).
\end{equation}
Therefore the first correction is proportional to the solution at leading order
and is given by
\begin{equation}
    \label{eqn:C1_solution}
    \C1(X) = \frac{1}{2}q_0\C0(0)\C0(X),
\end{equation}
finally giving $C=\Ch(x)+\tfrac{1}{2}\ep q_0 \Ch(0)\Ch(x)+\ep^2 q_0 \Ch(x) f(x')+\bo{\ep^3}$, missing one term in (\ref{eqn:C_a}) and the $\bo{\ep^2}$ term in (\ref{eqn:C_aa}).

\section{Unit cell integration identities}
\label{app:identities}
For $j = 1,2,\dotsc,N$,
\begin{equation}
    \label{eqn:unit_cell_poly}
    \int_{j-1/2}^{j+1/2} (y-j)^n [\delta(y-j) - 1] \dd y =
    \begin{cases}
        0, & n=0,1,\\
        -\tfrac{1}{12}, & n=2.
    \end{cases}
\end{equation}
For $k-1/2 < x < k+1/2$ (\ie $k = \floor{x+1/2}$),
\begin{subequations}
    \begin{equation}
        \int_{k-1/2}^x [\delta(y-k) - 1] \dd y = H(x-k) - x + k - \tfrac{1}{2},
    \end{equation}
    \begin{equation}
        \int_x^{k+1/2} [\delta(y-k) - 1] \dd y = H(k-x) + x - k - \tfrac{1}{2},
    \end{equation}
\end{subequations}

\begin{subequations}
    \begin{equation}
        \int_{k-1/2}^x (y-x) [\delta(y-k) - 1] \dd y =
        (k-x)H(x-k) + \tfrac{1}{2}(x-k+\tfrac{1}{2})^2,
    \end{equation}
    \begin{equation}
        \int_x^{k+1/2} (y-x) [\delta(y-k) - 1] \dd y =
        (k-x)H(k-x) - \tfrac{1}{2}(x-k-\tfrac{1}{2})^2,
    \end{equation}
\end{subequations}

\begin{subequations}
    \begin{equation}
        \int_{k-1/2}^x (y-x)^2 [\delta(y-k) - 1] \dd y =
        (k-x)^2 H(x-k) - \tfrac{1}{3}(x-k+\tfrac{1}{2})^3,
    \end{equation}
    \begin{equation}
        \int_x^{k+1/2} (y-x)^2 [\delta(y-k) - 1] \dd y =
        (k-x)^2 H(k-x) + \tfrac{1}{3}(x-k-\tfrac{1}{2})^3.
    \end{equation}
\end{subequations}

\section{Approximating sums with integrals}
\label{app:approx_sums_int}
Let $f(y)$ be a smooth function, except possibly at $y=k$, with $f(y) =
\bo{\ep^a}$ as $\ep \to 0$ for some $a \in \mathbb{R}$. Additionally assume
that successive derivatives of $f(y)$ fall in magnitude by a factor of
$\ep$; \ie $\diff{^n\! f}{y^n} = \bo{\ep^{a + n}}$. We decompose the integral
$\int_0^{x^-} f(y) \dd y$ into contributions from unit cells, centred at $y=j$,
$j = 1,2,\dotsc,k-1$, a half-cell from $y=0$ and the remaining interval
$[k-\tfrac{1}{2},x]$; we split the integral $\int_{x^+}^\epm f(y) \dd y$
similarly. Then
\begin{equation}
    \label{eqn:split_integrals}
    \begin{aligned}
        \int_0^{x^-} f(y) \dd y &= \int_0^{1/2} f(y) \dd y + \sum_{j=1}^{k-1}
        \int_{j-1/2}^{j+1/2} f(y) \dd y + \int_{k-1/2}^{x^-} f(y) \dd y\\
        \int_{x^+}^\epm f(y) \dd y &= \int_{x^+}^{k+1/2} f(y) \dd y +
        \sum_{j=k+1}^N \int_{j-1/2}^{j+1/2} f(y) \dd y +
        \int_{\epm-1/2}^\epm f(y) \dd y
    \end{aligned}
\end{equation}
Taylor expanding around $y=0$ and $y=\epm$ for the half-cells, around the
centre of each unit cell and $y=x^\pm$ for the intervals from and up to $x$,
\begin{equation}
    \label{eqn:taylor_expansions}
    \begin{aligned}
        f(y) &= f(0) + y f_y(0) + \tfrac{1}{2}y^2 f_{yy}(0) + \dotsb,\\
        f(y) &= f(j) + (y-j)f_y(j) + \tfrac{1}{2}(y-j)^2 f_{yy}(j) + \dotsb,\\
        f(y) &= f(x^\pm) + (y-x)f_y(x^\pm) + \tfrac{1}{2}(y-x)^2
        f_{yy}(x^\pm) + \dotsb,\\
        f(y) &= f(\epm) + (y-\epm) f_y(\epm) + \tfrac{1}{2}(y-\epm)^2
        f_{yy}(\epm) + \dotsb.
    \end{aligned}
\end{equation}
Integrating each term in \eqref{eqn:taylor_expansions},
\begin{equation}
    \label{eqn:integrated_taylor}
    \begin{aligned}
        \int_0^{1/2} f(y) \dd y &= \tfrac{1}{2} f(0) + \tfrac{1}{8}f_y(0) +
        \tfrac{1}{48}f_{yy}(0) + \dotsb,\\
        \int_{j-1/2}^{j+1/2} f(y) \dd y &= f(j) + \tfrac{1}{24}f_{yy}(j) +
            \dotsb,\\
        \int_{k-1/2}^{x^-} f(y) \dd y &= f(x^-)(x-k+\tfrac{1}{2}) -
        \tfrac{1}{2} f_y(x^-)(x-k+\tfrac{1}{2})^2 + \tfrac{1}{6}
        f_{yy}(x^-)(x-k+\tfrac{1}{2})^3 + \dotsb,\\
        \int_{x^+}^{k+1/2} f(y) \dd y &= f(x^+)(k-x+\tfrac{1}{2}) +
        \tfrac{1}{2} f_y(x^+)(k-x+\tfrac{1}{2})^2 + \tfrac{1}{6}
        f_{yy}(x^+)(k-x+\tfrac{1}{2})^3 + \dotsb,\\
        \int_{\epm-1/2}^{\epm} f(y) \dd y &= \tfrac{1}{2} f(\epm) -
        \tfrac{1}{8}f_y(\epm) + \tfrac{1}{48}f_{yy}(\epm) + \dotsb.
    \end{aligned}
\end{equation}
Using \eqref{eqn:integrated_taylor} in \eqref{eqn:split_integrals} and
rearranging, we have
\begin{equation}
    \begin{aligned}
        \sum_{j\neq k}f(j) = &\left(\int_0^{x^-} + \int_{x^+}^\epm\right)
        f(y) \dd y\\
        &- \tfrac{1}{2}[f(0) + f(\epm)] - [f(x^+)(k-x+\tfrac{1}{2}) +
        f(x^-)(x-k+\tfrac{1}{2})]\\
        &- \tfrac{1}{8}[f_y(0) - f_y(\epm)] - \tfrac{1}{24}\sum_{j\neq k}
        f_{yy}(j) - \tfrac{1}{2}[f_y(x^+)(k-x+\tfrac{1}{2})^2 -
        \tfrac{1}{2}f_y(x^-)(x-k+\tfrac{1}{2})^2]\\
        &+ \dotsb
    \end{aligned}
    \label{eqn:sum_to_int}
\end{equation}

\section{Averaging unit-cell fluctuations}
\label{app:shifted_envelope}

We seek $F(x)=\E{ f(x-k-\phi)}$ where $k=\floor{x+\tfrac{1}{2}}$ and $f(x)=-\tfrac{1}{2}(x^2 - |x| +
\tfrac{1}{6})$ in $|x|\leq \tfrac{1}{2}$ (and is zero otherwise), with $\phi\sim
\mathcal{N}(0,\sigma^2)$, with $\sigma \ll 1$. We restrict attention to the cell
$k=0$. Where $f$ varies smoothly with respect to $x$ (away from $x=0$),
$F(x)=f(x) + \bo{\sigma^2}$. However the spike in $f$ near $x=0$ is smoothed
more dramatically. Let $g(\phi)=(2\pi \sigma^2)^{-1/2}
\exp(-\phi^2/(2\sigma^2))$ be the probability density of the shift $\phi$. Then
for $x$ in a region of width $\bo{\sigma}$ near the origin,
\begin{equation}
    \begin{aligned}
        F(x) &=  \int_{-\infty}^{\infty}\left(-\frac{1}{12}+\frac{1}{2}\vert x-\phi \vert +\dots\right)g(\phi) \dd \phi +\bo{\sigma^2} \\
             &= -\frac{1}{12} + \int_{-\infty}^x \frac{1}{2} (x-\phi)g(\phi) \dd \phi + \int_{x}^{\infty} \frac{1}{2} (\phi-x) g(\phi) \dd \phi +\dots \\
             &= -\frac{1}{12} + \frac{\sigma}{\sqrt{2\pi}} \exp\left(-\frac{x^2}{2\sigma^2}\right) + \frac{x}{2} \erf\left(\frac{x}{\sigma\sqrt{2}}\right) +\dots
    \end{aligned}
\end{equation}
The outer limit of this approximation, for $x\gg \sigma$, gives $F\approx
-\tfrac{1}{12}+\tfrac{1}{2} \vert x\vert+\bo{\sigma^2}$, matching with the region
where $f$ varies smoothly.  For $x\ll \sigma$, 
\begin{equation}
    F\approx -\frac{1}{12}+ \frac{\sigma}{\sqrt{2\pi}} + \frac{1}{2\sigma}\frac{x^2} {\sqrt{2\pi}}+\bo{\sigma^2}.
\end{equation}  
Thus small normal perturbations of sink locations reduce the amplitude of
unit-cell fluctuations from $(-\tfrac{1}{12}, \tfrac{1}{24})$ (the range of $f$)
to $(-\tfrac{1}{12}+\sigma/\sqrt{2\pi},\tfrac{1}{24})$ (the range of $F$, with
error $\bo{\sigma^2}$). Stronger disorder suppresses the range completely: when
$\phi\sim\mathcal{U}(-\tfrac{1}{2},\tfrac{1}{2})$, 
\begin{equation}
    F(x)=\int_{-1/2}^{1/2} \sum_k f(x-k-\phi) \dd \phi = \int_{x-1/2}^{x+1/2} \sum_k f(u-k)\dd u = 0.
\end{equation}

\end{document}

%% file: preamble.tex
\usepackage{bm}
\usepackage{mathtools}
\usepackage{amsfonts}
\usepackage{color}
\usepackage{stmaryrd}
\usepackage{ulem}
\usepackage{grffile}

\newcommand{\diff}[2]{\frac{\mathrm{d} #1}{\mathrm{d} #2}}
\newcommand{\pdiff}[2]{\frac{\partial #1}{\partial #2}}

\newcommand{\ie}{i.e.\ }
\newcommand{\eg}{e.g.\ }

\DeclareMathOperator*{\Var}{Var}
\DeclareMathOperator*{\Cov}{Cov}

\newcommand{\E}[1]{\left\langle #1 \right\rangle}

\newcommand{\var}[1]{\Var\left(#1\right)}
\newcommand{\cov}[2]{\Cov\left(#1,#2\right)}

\newcommand{\tcov}[1]{\Cov\nolimits^\text{T}\left(#1\right)}

\newcommand{\dd}[1]{\,\mathrm{d} #1}

\newcommand{\bo}[1]{O(#1)}

\DeclareMathOperator{\pe}{Pe}
\DeclareMathOperator{\da}{Da}

\newcommand{\ep}{\varepsilon}
\newcommand{\epm}{{\varepsilon^{-1}}}

\newcommand{\C}[1]{C^{(#1)}}

\DeclarePairedDelimiter\floor{\lfloor}{\rfloor}

\DeclarePairedDelimiter{\jump}{\llbracket}{\rrbracket}

\newcommand{\ti}[1]{\widetilde{#1}}

\newcommand{\lx}{\mathcal{L}}
\newcommand{\bx}{\mathcal{B}}

\newcommand{\lapp}{\mathcal{L}_x}
\newcommand{\bapp}{\mathcal{B}_x}

\newcommand{\Ch}{C_\text{H}}
\newcommand{\Chs}[1]{C_{\text{H}#1}}

\newcommand{\gch}{(G\Ch)}
\newcommand{\gpch}{(G^+\Ch)}
\newcommand{\gmch}{(G^-\Ch)}

\newcommand{\rc}{r_{\text{c}}}

\DeclareMathOperator{\erf}{erf}

%% file: paper_2_110918_arxiv.bbl
\begin{thebibliography}{}

\bibitem[Ahsanullah et~al., 2013]{ahsanullah2013introduction}
Ahsanullah, M., Nevzorov, V.~B. {\&} Shakil, M. (2013) {\em An Introduction to
  Order Statistics}.
Springer.

\bibitem[Bal, 2011]{bal2011}
Bal, G. (2011)  Convergence to homogenized or stochastic partial differential
  equations. {\em App. Math. Res. eXpress}, \textbf{2011}(2), 215--241.

\bibitem[Bruna \& Chapman, 2015]{bruna2015diffusion}
Bruna, M. {\&} Chapman, S.~J. (2015)  Diffusion in spatially varying porous
  media. {\em SIAM J. Appl. Math.}, \textbf{75}(4), 1648--1674.

\bibitem[Burridge \& Keller, 1981]{burridgekeller1981}
Burridge, R. {\&} Keller, J.~B. (1981)  Poroelasticity equations derived from
  microstructure. {\em J. Acoust. Soc. Am.}, \textbf{70}(4), 1140--1146.

\bibitem[Chernyavsky et~al., 2010]{chernyavsky2010mathematical}
Chernyavsky, I., Jensen, O. {\&} Leach, L. (2010)  A mathematical model of
  intervillous blood flow in the human placentone. {\em Placenta},
  \textbf{31}(1), 44--52.

\bibitem[Chernyavsky et~al., 2012]{chernyavsky2012characterizing}
Chernyavsky, I.~L., Dryden, I.~L. {\&} Jensen, O.~E. (2012)  Characterizing the
  multiscale structure of fluctuations of transported quantities in a
  disordered medium. {\em IMA J. Appl. Math.}, \textbf{77}(5), 697--725.

\bibitem[Chernyavsky et~al., 2011]{chernyavsky2011transport}
Chernyavsky, I.~L., Leach, L., Dryden, I.~L. {\&} Jensen, O.~E. (2011)
  Transport in the placenta: homogenizing haemodynamics in a disordered medium.
  {\em Phil. Trans. R. Soc. A}, \textbf{369}(1954), 4162--4182.

\bibitem[Chunsheng, 1992]{chunsheng1992moments}
Chunsheng, M. (1992)  Moments of functions of order statistics. {\em Statistics
  \& Prob. Lett.}, \textbf{15}(1), 57--62.

\bibitem[Cushman et~al., 2002]{Cushman2002}
Cushman, J.~H., Bennethum, L.~S. {\&} Hu, B.~X. (2002)  A primer on upscaling
  tools for porous media. {\em Adv. Water Res.}, \textbf{25}(8), 1043--1067.

\bibitem[David \& Nagaraja, 2003]{david2003order}
David, H.~A. {\&} Nagaraja, H.~N. (2003) {\em Order Statistics}.
John Wiley \& Sons, Inc., 3 edition.

\bibitem[Davit et~al., 2013]{davit2013homogenization}
Davit, Y., Bell, C.~G., Byrne, H.~M., Chapman, L.~A., Kimpton, L.~S., Lang,
  G.~E., Leonard, K.~H., Oliver, J.~M., Pearson, N.~C., Shipley, R.~J.  et~al.
  (2013)  Homogenization via formal multiscale asymptotics and volume
  averaging: How do the two techniques compare?. {\em Adv. Water Res.},
  \textbf{62}, 178--206.

\bibitem[Erian et~al., 1977]{erian1977maternal}
Erian, F., Corrsin, S. {\&} Davis, S. (1977)  Maternal, placental blood flow: a
  model with velocity-dependent permeability. {\em J. Biomech.},
  \textbf{10}(11-12), 807--814.

\bibitem[{Erlich} et~al., 2018]{erlich2018}
{Erlich}, A., {Pearce}, P., {Plitman Mayo}, R., {Jensen}, O.~E. {\&}
  {Chernyavsky}, I.~L. (2018)  {Physical and geometric determinants of
  transport in feto-placental microvascular networks}. {\em arXiv:1809.00749}.

\bibitem[Heitzinger \& Ringhofer, 2014]{heitzinger2014}
Heitzinger, C. {\&} Ringhofer, C. (2014)  Multiscale modeling of fluctuations
  in stochastic elliptic PDE models of nanosensors. {\em Commun. Math. Sci},
  \textbf{12}(3), 401--421.

\bibitem[Jensen \& Chernyavsky, 2019]{jensenchernyavsky2019}
Jensen, O.~E. {\&} Chernyavsky, I.~L. (2019)  Blood Flow and Transport in the
  Human Placenta. {\em Ann. Rev. Fluid Mech.}, \textbf{51}, 25--47.

\bibitem[Le~Bris, 2014]{lebris2014}
Le~Bris, C. (2014)  Homogenization theory and multiscale numerical approaches
  for disordered media: some recent contributions. {\em ESAIM: Proc. Surveys},
  \textbf{45}, 18--31.

\bibitem[Pavliotis \& Stuart, 2008]{pavliotisstuart2008}
Pavliotis, G. {\&} Stuart, A. (2008) {\em Multiscale methods: averaging and
  homogenization}.
Springer Science \& Business Media.

\bibitem[Rubinstein \& Torquato, 1989]{rubinsteintorquato1989}
Rubinstein, J. {\&} Torquato, S. (1989)  Flow in random porous media:
  mathematical formulation, variational principles, and rigorous bounds. {\em
  J. Fluid Mech.}, \textbf{206}, 25--46.

\bibitem[Russell et~al., 2016]{russell2016stochastic}
Russell, M.~J., Jensen, O.~E. {\&} Galla, T. (2016)  Stochastic transport in
  the presence of spatial disorder: Fluctuation-induced corrections to
  homogenization. {\em Phys. Rev. E}, \textbf{94}(4), 042121.

\bibitem[Schmuck \& Bazant, 2015]{schmuck2015}
Schmuck, M. {\&} Bazant, M.~Z. (2015)  Homogenization of the
  Poisson--Nernst--Planck equations for ion transport in charged porous media.
  {\em SIAM J. Appl. Math.}, \textbf{75}(3), 1369--1401.

\bibitem[Wood \& Vald{\'e}s-Parada, 2013]{wood2013}
Wood, B.~D. {\&} Vald{\'e}s-Parada, F.~J. (2013)  Volume averaging: Local and
  nonlocal closures using a Green's function approach. {\em Adv. Water Res.},
  \textbf{51}, 139--167.

\end{thebibliography}
